%% file: main.tex
\documentclass[10pt,twocolumn,letterpaper]{article}

\usepackage{cvpr}              
\usepackage{arydshln}


\definecolor{cvprblue}{rgb}{0.21,0.49,0.74}
\usepackage[pagebackref,breaklinks,colorlinks,allcolors=cvprblue]{hyperref}

\usepackage{comment}
\usepackage{multirow}

\title{ELiC: Efficient LiDAR Geometry Compression via Cross-Bit-depth Feature Propagation and Bag-of-Encoders}

\author{
Junsik Kim \quad Gun Bang \quad Soowoong Kim \\
Electronics and Telecommunications Research Institute (ETRI) \\
{\tt\small \{junsikkim, gbang, soowoong.kim\}@etri.re.kr}
}

\begin{document}
\maketitle
\input{sec/0_abstract}    
\input{sec/1_intro}

\input{sec/2_related}

\input{sec/3_method}

\input{sec/4_experiments}
\input{sec/5_conclusion}
{
    \small
    \bibliographystyle{ieeenat_fullname}
    \bibliography{main}
}

\input{sec/X_suppl}

\end{document}

%% file: sec/0_abstract.tex
\begin{abstract}

Hierarchical LiDAR geometry compression encodes voxel occupancies from low to high bit-depths, yet prior methods treat each depth independently and re-estimate local context from coordinates at every level, limiting compression efficiency. 
We present ELiC, a real-time framework that combines cross-bit-depth feature propagation, a Bag-of-Encoders (BoE) selection scheme, and a Morton-order-preserving hierarchy. 
Cross-bit-depth propagation reuses features extracted at denser, lower depths to support prediction at sparser, higher depths. 
BoE selects, per depth, the most suitable coding network from a small pool, adapting capacity to observed occupancy statistics without training a separate model for each level. 
The Morton hierarchy maintains global Z-order across depth transitions, eliminating per-level sorting and reducing latency. 
Together these components improve entropy modeling and computation efficiency, yielding state-of-the-art compression at real-time throughput on Ford and SemanticKITTI.
Code and pretrained models are available at \url{https://github.com/moolgom/ELiCv1}.

\end{abstract}

%% file: sec/1_intro.tex
\section{Introduction}
\label{sec:intro}

\begin{figure}[t]
  \centering
  \includegraphics[width=0.95\linewidth]{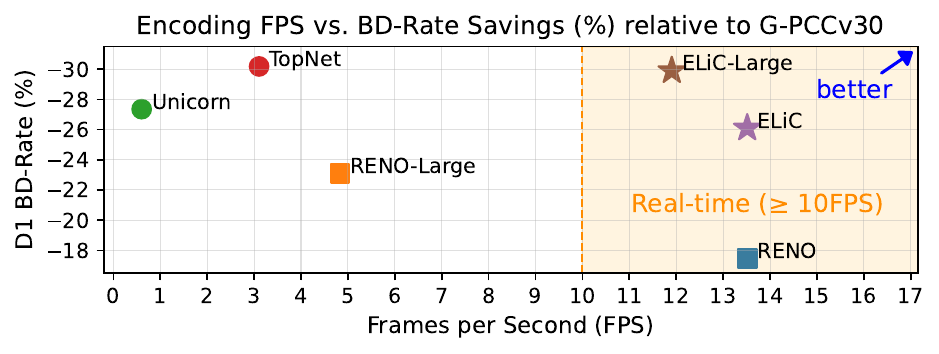}
  \includegraphics[width=0.95\linewidth]{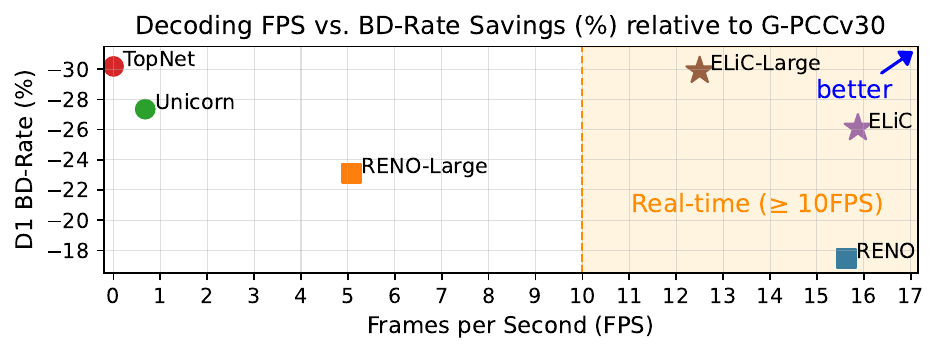}
  \caption{
    Average encoding and decoding FPS for 12-bit LiDAR geometry compression, and BD-Rate savings relative to G-PCC v30 across 16-, 15-, 14-, 13-, and 12-bit LiDAR geometries. Our models (ELiC and ELiC-Large) achieve real-time compression at 12-bit input (LiDAR capture speed ${\ge}10$ FPS) while maintaining compression efficiency comparable to state-of-the-art methods.
    Results are reported on Ford and SemanticKITTI datasets.
  }
  \label{fig:teaser}
\end{figure}

LiDAR is widely used in autonomous driving, robotics, V2X, digital twins, and large-scale 3D mapping because it captures precise, large-scale geometry \cite{Yang_2024_CVPR, mounier2024lidar, gao2024damage, zimmer2024tumtraf, hu2024bim, Lin_2025_CVPR}. 
Raw LiDAR point clouds are massive, creating storage and transmission bottlenecks. 
Robots and vehicles must run perception under tight compute and power budgets, often in real time. 
Offloading to edge or cloud helps, but the uplink must still carry point clouds under strict bandwidth and latency constraints. 
Hence, real-time LiDAR geometry compression is needed to cut bitrate and delay while preserving the structural fidelity required by downstream tasks.

Several recent sparse tensor-based point cloud geometry compression techniques \cite{SparsePCGC, Unicorn, RENO, UniPCGC} adopt a bit-depth-level-independent design with a single shared sparse convolution-based coding network. 
These methods construct an octree-style hierarchy by progressively coarsening the input point cloud from its original bit-depth. 
At each level, they take the voxel coordinates corresponding to that bit-depth as input, independently predict the occupancy probabilities of the next-level voxels, and entropy-code the voxel-occupancy labels according to their likelihoods.
Compared with pipelines that downsample multiple times, compress latent features, and then mirror the process with upsamplings \cite{wang2021multiscale, zhang2023yoga, qi2024variable, akhtar2024inter}, these methods are more flexible and directly operate at variable input bit-depth levels of the point cloud geometry. 
However, the independence across levels remains a core weakness: at each level, the model must re-derive spatial context that was already available at earlier, previous denser levels, due to its reliance solely on the coordinates at the current bit-depth level.

\begin{figure}[t!]
  \centering
  \includegraphics[width=0.7\linewidth]{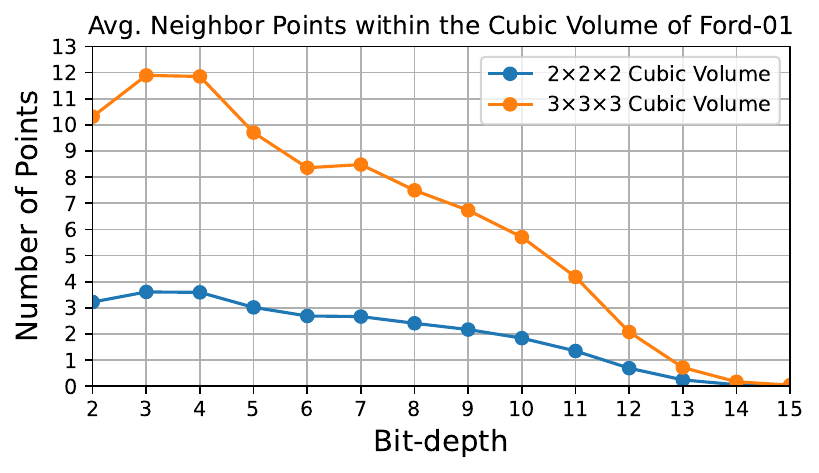}
  \caption{Average number of neighboring points by coordinate resolution (bit-depth) in the Ford-01 sequence (1,500 frames).}
  \label{fig:npcounts}
\end{figure}

For LiDAR geometry compression, this drawback is particularly consequential. 
Fig.~\ref{fig:npcounts} shows the average number of neighboring points within $2{\times}2{\times}2$ and $3{\times}3{\times}3$ cubic volumes versus each bit-depth level on the Ford-01 sequence \cite{CTC}. 
At the 13--15-bit levels, the average count is at most one neighbor, indicating insufficient local geometric information. 
Because sparse convolutional networks typically operate with $2{\times}2{\times}2$ or $3{\times}3{\times}3$ kernels, such low neighbor counts leave their receptive fields nearly empty, forcing the model to reconstruct spatial context from minimal evidence.
Consequently, even when using 14-bit coordinates\textemdash or anything one bit lower\textemdash to infer 15-bit occupancy, the available signal remains severely limited.

In addition, Fig.~\ref{fig:npcounts} indicates that the point density distribution varies significantly across bit-depth levels.
At lower levels, the points are densely located, whereas they become increasingly sparse as the bit-depth increases.
This density shift alters neighbor statistics, activation rates, and effective receptive-field usage.
Therefore, employing a single shared coding network model makes it difficult to achieve optimal compression efficiency across bit-depth levels.

We address these issues by extending the lightweight LiDAR geometry compression model RENO~\cite{RENO}.
While RENO extracts contextual information only from one bit-depth level lower than the current coordinates, our approach enhances this process by propagating and reusing feature information from multiple lower bit-depth levels to higher ones.
As these features accumulate from the lowest level ($b{=}2$) upward, the model progressively retains and transfers geometric context to increasingly sparse bit-depth levels, thereby supporting occupancy prediction even when immediate neighbors are scarce.

Next, we replace the level-independent single model with a Bag-of-Encoders (BoE) framework.
The BoE is a compact pool of coding networks that share the same architecture but have distinct network parameters.
At each bit-depth level, we compute a lightweight descriptor of the next-level voxel-occupancy distribution, use this descriptor to select the most suitable network from the pool, and encode that level using the selected model.
The indices of the selected model are transmitted to ensure identical behavior between the encoder and decoder across all bit-depth levels.
Even when different models are selected for consecutive levels, cross-bit-depth feature propagation remains intact.
This design enables the model to adapt to the varying density distribution across bit-depth levels without training a separate network for each level.

Separately, we adopt a Morton-order--preserving hierarchy that maintains a global Z-order~\cite{morton1966computer} across bit-depth transitions, eliminating per-level sorting overhead and reducing encoding and decoding runtime.
Together, these three components—Morton-order--preserving hierarchy, cross-bit-depth feature propagation, and BoE selection—constitute our model, ELiC, which achieves higher real-time throughput and superior compression efficiency than recent LiDAR geometry compression models.

%% file: sec/2_related.tex
\section{Related Works}
\label{sec:related}

\subsection{Conventional and Standardized Approaches}
Draco~\cite{draco} employs a recursive K-D tree partitioning along the longest axis and applies arithmetic coding to the k-d tree statistics, enabling fast real-time geometry compression.
G-PCC~\cite{GPCC} operates directly in 3D space using an octree structure and compresses it with context-adaptive arithmetic coding.
V-PCC~\cite{VPCC} maps 3D to 2D patches for video coding, effective on dynamic data but inefficient on sparse LiDAR due to many empty pixels from projection/packing.

\subsection{Learning-based Approaches}

By representation and processing strategy, learning-based methods fall into three groups: 2D range image-based, tree-based, and sparse tensor-based.

\noindent \textbf{2D range image-based methods} project LiDAR geometry onto 2D range maps, utilizing the grid structure to apply 2D learning techniques for efficient feature extraction and entropy modeling. 
RIDDLE~\cite{zhou2022riddle} introduces a PixelCNN~\cite{van2016pixel}-like predictor that sequentially estimates per-pixel attributes in raster order from decoded range-image patches. 
BIRD-PCC~\cite{liu2023bird} adopts bidirectional predictive coding with invalid-pixel masking and deep residual coding to exploit spatial redundancy. 
Zhao et al.~\cite{zhao2022real} improve efficiency via scanline-based predictive coding and adaptive residual entropy coding on range images.
While range-image methods are computationally efficient and leverage mature image processing techniques, the projection suffers from occlusion: points closer to the sensor remain visible, whereas farther points are discarded, leading to degraded fidelity in occluded regions~\cite{wu2021detailed}.

\noindent \textbf{Tree-based methods} utilize hierarchical structures to efficiently capture the multiscale nature of point clouds. 
OctAttention~\cite{OctAttention} introduces an attention mechanism into the octree domain to model long-range spatial dependencies, leading to improved rate-distortion (R-D) performance. 
ECM-OPCC~\cite{ECMOPCC} employs dual transformer networks combined with a segment-constrained multi-group strategy to balance modeling capacity and decoding efficiency. 
VoxelContext-Net~\cite{VoxelContextNet} further enhances performance by incorporating voxel-level context into octree structures. 
TopNet~\cite{Wang_2025_CVPR} employs a hierarchical transformer to predict per-node 8-way occupancies on the octree, achieving strong R-D performance. 
Although tree-based methods achieve strong R-D performance, their node-level sequential reconstruction and heavy context modeling typically lead to slow decoding.


\noindent \textbf{Sparse tensor-based methods} quantize point clouds into voxel grids and process them using 3D sparse convolutional architectures.
SparsePCGC~\cite{SparsePCGC} models multiscale voxel dependencies to improve geometry entropy coding.
Unicorn~\cite{Unicorn} employs a unified multiscale conditional model that achieves strong rate-distortion performance by a higher computational cost.
UniPCGC~\cite{UniPCGC} provides a single framework supporting variable rate and complexity with competitive rate-distortion efficiency.
RENO~\cite{RENO} adopts a two-stage occupancy coding with lightweight sparse networks to reach real-time throughput.
However, at higher bit-depth levels where LiDAR geometry becomes extremely sparse, the fixed receptive fields of sparse convolutions often fail to capture sufficient spatial context, thereby limiting coding efficiency and reconstruction accuracy.


%% file: sec/3_method.tex
\begin{figure*}[!th]
  \centering
  \includegraphics[width=0.9\linewidth, trim=0 9 0 13, clip]{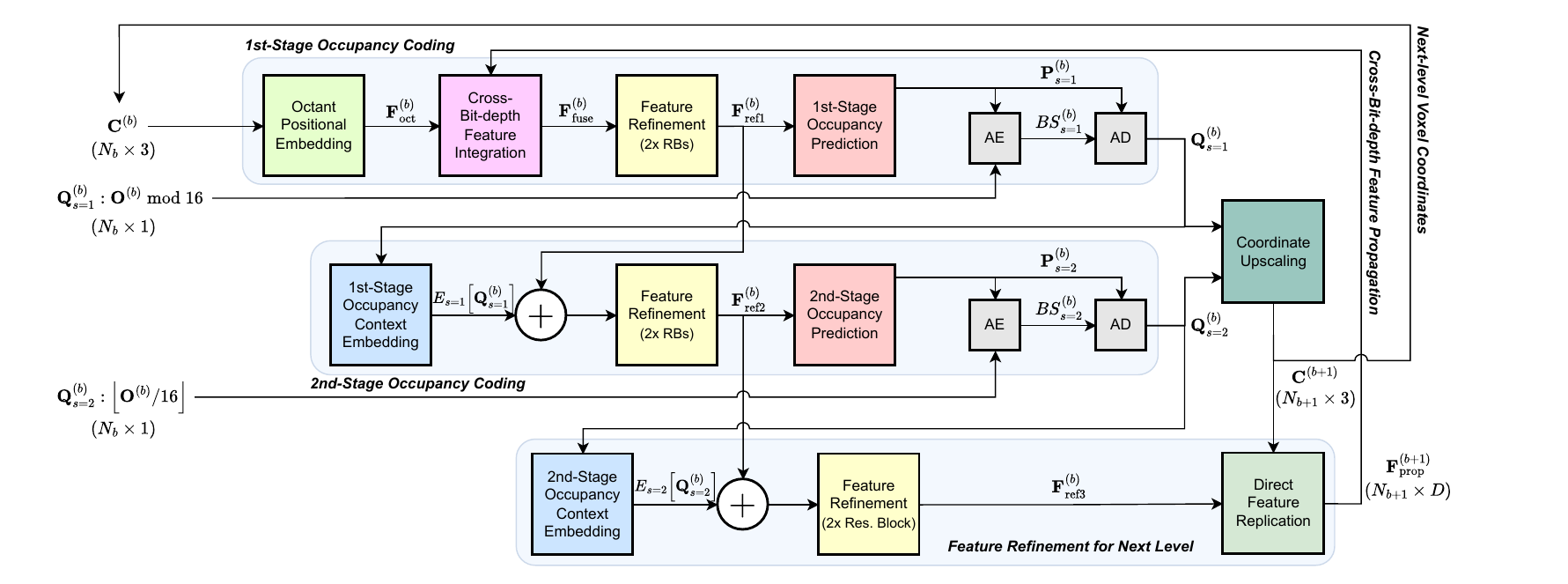}
  \caption{Architecture of the coding network in ELiC. The diagram shows encoding and decoding pipelines together. For the separated pipelines, see Fig.~\ref{fig:pipeline} in the supplementary material.}
  \label{fig:codingnet}
\end{figure*}

\section{Progressive Coding Preliminaries}

Given an input point cloud quantized to bit-depth $B$, our LiDAR geometry compression framework performs progressive encoding and decoding over bit-depth levels from $b{=}2$ to $b{=}B{-}1$. 
At each level $b$, the input is an $N_b{\times}3$ integer coordinate matrix $\mathbf{C}^{(b)}\!\in\!\mathbb{Z}^{N_b\times3}$, representing voxel coordinates at level $b$. For each input point (i.e., occupied parent voxel), the encoder predicts the $(b{+}1)$-level occupancy pattern of its eight child voxels and arithmetic-codes the true 8-bit symbol; the decoder, in turn, arithmetic-decodes the bitstream.
This is equivalent to octree coding with per-node occupancy prediction, where the supervision target is an octant label $\mathbf{O}^{(b)}\!\in\!\{0,\ldots,255\}^{N_b}$.

Following RENO~\cite{RENO}, our model does not predict the octant label  $\mathbf{O}^{(b)}$ in a single step.
Instead, we adopt the same two-stage coding strategy, denoted by the stage index $s\!\in\!\{1, 2\}$: the first-stage ($s{=}1$) predicts the lower quadrant label, and the second-stage ($s{=}2$) predicts the upper quadrant label conditioned on the first-stage output. 
This factorization reduces the symbol space from $2^8$ to $2^4$ per stage, simplifying probability modeling and improving entropy coding efficiency.

Formally, the stage-wise quadrant label is defined as
\begin{equation}
\label{eq:two_stage_labels}
\mathbf{Q}_{s}^{(b)} = 
\begin{cases}
\mathbf{O}^{(b)} \bmod 16 & (s=1), \\
\left\lfloor \mathbf{O}^{(b)}/{16} \right\rfloor & (s=2),
\end{cases}
\end{equation}
where $\bmod$ and $\lfloor\cdot\rfloor$ are applied element-wise,
so that $\mathbf{Q}_{s}^{(b)} \in \{0,\ldots,15\}^{N_b}$.
The original octant label can be reconstructed as
\begin{equation}
\label{eq:two_stage_reconstruct}
\mathbf{O}^{(b)} \;=\; 16\,\mathbf{Q}_{s=2}^{(b)} \;+\; \mathbf{Q}_{s=1}^{(b)}.
\end{equation}

At each level $b$, the encoder takes $\mathbf{C}^{(b)}$ and the corresponding target $\mathbf{O}^{(b)}$ as input and produces the stage-wise bitstreams $\{BS_{s}^{(b)}\}_{s=1,2}$.
Conversely, the decoder takes $\mathbf{C}^{(b)}$ and $\{BS_{s}^{(b)}\}_{s=1,2}$ as input and reconstructs the $\mathbf{O}^{(b)}$.
From the $\mathbf{O}^{(b)}$, the next-level coordinate $\mathbf{C}^{(b{+}1)}$ is derived by enumerating the positions of the occupied child voxels.
This process is repeated for $b=2,\ldots,B{-}1$ until the target level-$B$ coordinates $\mathbf{C}^{(B)}$ are reconstructed.

\section{Proposed Method}
\label{sec:method}

We propose ELiC, which preserves RENO's lightweight, real-time design while improving compression efficiency. 
Rather than introducing heavy, state-of-the-art modules, we optimize information flow and adaptively select the coding network within a compact pipeline


ELiC is built upon three core ideas.
\textbf{(1) Morton-order-preserving hierarchy:} We establish a global 3D Morton~\cite{morton1966computer} for all coordinates and maintain it across bit-depth transitions via integer down/up-scaling, eliminating any re-sorting during traversal.
\textbf{(2) Cross-bit-depth feature propagation:} Latent features extracted at lower bit-depth levels are propagated upward and fused with current-level features, reducing uncertainty in occupancy prediction and improving entropy coding efficiency.
\textbf{(3) Bag-of-Encoders (BoE):} To handle varying occupancy distributions across bit-depths, we construct a compact pool of coding networks with a shared architecture but distinct parameters, improving adaptability without increasing complexity.

\subsection{Morton-Order-Preserving Hierarchy}
\label{sec:coords}


Prior methods \cite{SparsePCGC, UniPCGC, RENO} often enforce encoder-decoder consistency by repeatedly sorting coordinates and reordering features during hierarchy traversal, which incurs sorting overhead.
We instead use a 3D Morton hierarchy to make the coordinate-feature order hierarchically invariant, eliminating all sorting across bit-depth transitions.

We first sort the input coordinates $\mathbf{C}^{(B)}$ by their 3D Morton codes.
Subsequent downscaling is performed by integer division $\mathbf{C}^{(b-1)} = \lfloor \mathbf{C}^{(b)} / 2 \rfloor$, which preserves the Morton order at all coarser levels.
Repeating this process down to $b{=}2$ yields a hierarchy in which all $\mathbf{C}^{(b)}$ share the same global Morton sequence.
Consequently, points belonging to the same parent always occupy contiguous intervals, and each parent's order directly contains that of its children.
Based on this structure, we encode each parent's occupancy $\mathbf{O}^{(b)}$ as an 8-bit pattern representing the occupied octants.

During upscaling, each parent coordinate is doubled and expanded with the fixed set of octant offsets $\{\boldsymbol{\delta}_u\}_{u=0}^{7}\subset\{0,1\}^3$ (preordered in Morton octant order) to generate all candidate child coordinates:
\begin{equation}
\widetilde{\mathbf{C}}^{(b+1)}
= 2\big(\mathbf{C}^{(b)} \otimes \mathbf{1}_{8}\big)
+ \big(\mathbf{1}_{N_b}\otimes \{\boldsymbol{\delta}_u\}_{u=0}^{7}\big).
\end{equation}
Define the per-parent, per-octant mask $\mathbf{M}^{(b)}\in\{0,1\}^{N_b\times 8}$ by
\begin{equation}
\label{eq:occ-mask}
\begin{aligned}
\mathbf{M}^{(b)}(n,u) &= \Big\lfloor \mathbf{O}^{(b)}(n) / 2^{u} \Big\rfloor \bmod 2, \\
                      &\forall\, n\in\{1,\ldots,N_b\},\; u\in\{0,\ldots,7\}.
\end{aligned}
\end{equation}
Flattening $\mathbf{M}^{(b)}$ in row-major order yields a vector in $\{0,1\}^{8N_b}$, which we use to index $\widetilde{\mathbf{C}}^{(b+1)}$ and obtain the actual child coordinates:
\begin{equation}
\label{eq:child-select}
\mathbf{C}^{(b+1)}
= \widetilde{\mathbf{C}}^{(b+1)}\big|_{\mathbf{M}^{(b)}=1}.
\end{equation}
Because both the parent and octant enumerations follow the Morton rule, $\mathbf{C}^{(b+1)}$ automatically preserves global Morton order without any explicit re-sorting.
A single initial sort eliminates all subsequent re-sorting.
For more details, please refer to the supplementary material (Sec.~\ref{sec:zorder1} and~\ref{sec:zorder2}).

\begin{figure*}[!t]
  \centering
  \includegraphics[width=0.9\linewidth, trim=0 6 0 5, clip]{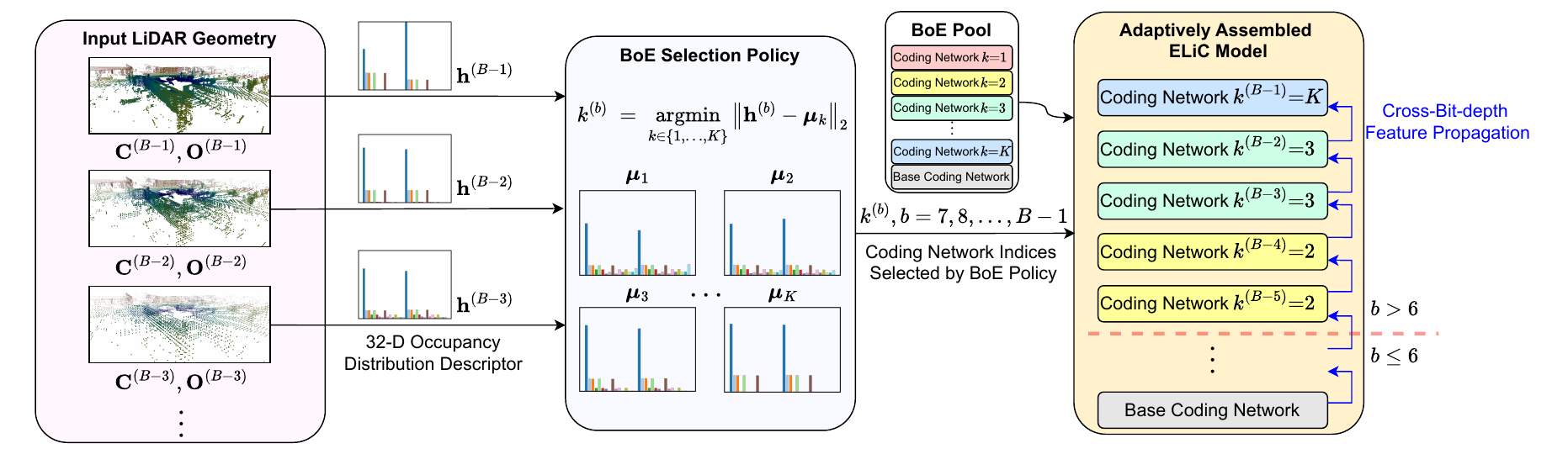}
  \caption{Concept diagram of the Bag-of-Encoders (BoE) coding strategy in ELiC. At each bit-depth level, ELiC selects a coding network from a BoE pool based on the occupancy distribution and adaptively assembles the model for LiDAR geometry compression.}
  \label{fig:BoE}
\end{figure*}


\subsection{Coding Network Architecture}

Fig.~\ref{fig:codingnet} illustrates the architecture of the proposed coding network.
At each bit-depth level $b$, given the voxel coordinates $\mathbf{C}^{(b)}$ and the propagated features from the previous level $\mathbf{F}_{\text{prop}}^{(b)}$ (carried over from $b{-}1$), the network predicts the stage-wise quadrant labels $\{\mathbf{Q}_{s}^{(b)}\}_{s=1,2}$.

\noindent \textbf{Octant Positional Embedding:}
This module provides a deterministic local positional prior by embedding the $2{\times}2{\times}2$  sub-voxel (octant) parity.
The embedding serves as the initial feature for coding of the current level $b$ and is fused with the cross-bit-depth feature propagated from lower levels.

Given an integer coordinate $\mathbf{c}=(x,y,z)\in\mathbf{C}^{(b)}$, define the octant parity
$\mathbf{p}=\big(x\bmod 2,\; y\bmod 2,\; z\bmod 2\big)\in\{0,1\}^3$ and its index
$i=\mathbf{p}\cdot[1,2,4]^\top\in\{0,\ldots,7\}$.
With an embedding table $E_{\text{oct}}\in\mathbb{R}^{8\times D}$, the octant feature is
$\mathbf{f}_{\text{oct}}=E_{\text{oct}}[i]\in\mathbb{R}^{D}$.

\noindent \textbf{Cross-Bit-depth Feature Integration:}
Unlike conventional level-independent occupancy coding methods~\cite{SparsePCGC, Unicorn, UniPCGC, RENO}, ELiC propagates and reuses features from lower occupancy coding levels. 
This propagation is progressively accumulated through learned blending weights. 
As bit-depth level increases and spatial coordinates become sparser, dense contextual information from lower levels is carried forward, which improves occupancy coding at highly sparse upper levels.
For $b{=}2$, no prior features exist, so fusion is bypassed and the octant feature is used directly.

Given the octant feature $\mathbf{F}_{\text{oct}}^{(b)}\in\mathbb{R}^{N_{b}{\times}D}$ and the propagated cross-bit-depth feature $\mathbf{F}_{\text{prop}}^{(b)}\in\mathbb{R}^{N_{b}{\times}D}$, we form a richer multi-scale representation at bit-depth level $b$ by fusing them via channel-wise gated fusion:
\begin{equation}
\label{eq:cs-fuse}
\begin{aligned}
\relax[\mathbf{w}_c,\; \mathbf{w}_p] &= \operatorname{softmax}(\mathbf{W}),\\
\mathbf{F}_{\text{fuse}}^{(b)} &= \mathbf{w}_c \odot \mathbf{F}_{\text{oct}}^{(b)} \;+\; \mathbf{w}_p \odot \mathbf{F}_{\text{prop}}^{(b)}.
\end{aligned}
\end{equation}
Here, $\mathbf{W}\in\mathbb{R}^{2\times D}$ is a learnable parameter that controls the channel-wise blending ratio between the two features. The softmax is applied per channel.
In the BoE, each coding network in the pool maintains its own gating parameter $\mathbf{W}_k$, allowing the fusion ratios to adapt to the occupancy distribution.

\noindent \textbf{Feature Refinement:} 
In the proposed model, feature refinement is performed by two consecutive plain residual blocks, each consisting of two $3{\times}3{\times}3$ sparse convolution layers.
At each bit-depth level, the refinement is applied three times: (1) before the first-stage occupancy prediction to stabilize the initial representation; 
(2) after embedding the first-stage occupancy context to strengthen features for the second-stage prediction; 
and (3) after embedding the second-stage occupancy context and before propagating to the next bit-depth level to consolidate information for cross-scale feature transfer. 

Our design further injects the second-stage occupancy context after occupancy coding and performs an additional refinement before propagating the features to the next bit-depth level, thereby distinguishing itself from prior methods~\cite{SparsePCGC, Unicorn, UniPCGC, RENO}, which compress each bit-depth level independently and terminate once that level is coded.

\noindent \textbf{Occupancy Prediction:} 
After the feature refinement in the first and second stages, predictions for the $\{\mathbf{Q}_{s}^{(b)}\}_{s=1,2}$ are produced, respectively. 
Each prediction head is consists of Linear$\rightarrow$ReLU$\rightarrow$Linear$\rightarrow$Softmax modules, yielding
$\mathbf{P}_{s}^{(b)} \in \mathbb{R}^{N_b \times 16}$,
which represents the 16-way symbol probability corresponding to the $\mathbf{Q}_{s}^{(b)}$.
During encoding, the $\mathbf{P}_{s}^{(b)}$ is used to arithmetic-encodes the $\mathbf{Q}_{s}^{(b)}$, while during decoding, this is used to arithmetic-decodes the $BS_{s}^{(b)}$.

\noindent \textbf{Occupancy Context Embedding:} 
After each coding stage, per-point features are enriched by adding a stage-specific context vector looked up from a dedicated $16{\times}D$ embedding table using the $\mathbf{Q}_{s}^{(b)}$.
Concretely, for each stage with the $\mathbf{Q}_{s}^{(b)}$, its own embedding table $E_{s}\!\in\!\mathbb{R}^{16\times D}$ is used, and the features are updated as
$\mathbf{F} \leftarrow \mathbf{F} \;+\; E_{s}\!\left[\mathbf{Q}_{s}^{(b)}\right]$, where $[\cdot]$ denotes per-point row indexing.
This deterministic embedding augments the representation without changing spatial resolution and prepares the feature for the subsequent refinement and propagation steps.

\noindent \textbf{Coordinate Upsacling:} 
After decoding level $b$, the next coordinates $\mathbf{C}^{(b+1)}$ are computed from the parent coordinates $\mathbf{C}^{(b)}$ and the octant label $\mathbf{O}^{(b)}$, following Eq.~\ref{eq:child-select}.

\noindent \textbf{Direct Feature Replication:} 
After the second-stage context has been added and the final refinement at bit-depth $b$ has completed, we replicate the refined feature $\mathbf{F}^{(b)}\in\mathbb{R}^{N_b\times D}$ to the occupied children using the occupancy mask  $\mathbf{M}^{(b)}$ (defined in Eq.~\ref{eq:occ-mask}), producing the propagated feature for the next depth:
\begin{equation}
\label{eq:feat-prop}
\mathbf{F}_{\text{prop}}^{(b+1)} \;=\; \big(\mathbf{F}^{(b)} \otimes \mathbf{1}_{8}\big)\big|_{\mathbf{M}^{(b)}=1}\;\in\mathbb{R}^{N_{b+1}\times D}.
\end{equation}



\subsection{Bag of Encoders}
\label{sec:BoE}

Occupancy distributions differ across bit-depth levels and scenes, making a single shared coding network suboptimal for all levels. 
Assigning a dedicated network to every bit-depth would improve adaptability but greatly increase the total model size. 
To balance adaptability and compactness, as illustrated in Fig.~\ref{fig:BoE}, we construct a compact pool of coding networks and, at each bit-depth level, select the most suitable one for encoding. 
Bit-depth levels $b{=}2$-$6$ are encoded using a base coding network, while higher levels are adaptively handled by selecting a network according to the occupancy distribution. 
The indices of the selected networks are signaled so that the decoder side applies the same models across bit-depths.

\noindent \textbf{Clustering-Based BoE Center Construction:} 
For a point set at level $b$, the distribution of the stage-wise quadarnt label $\mathbf{Q}_{s}^{(b)}$ is summarized as a 16-bin histogram $H_{s}\in\mathbb{R}_{\ge0}^{16}$. 
Concatenation followed by normalization (sum to one) yields a compact 32-D descriptor
\begin{equation}
\mathbf{h} \;=\; \frac{[\,H_{s{=}1};\,H_{s{=}2}\,]}{\mathbf{1}^\top[\,H_{s{=}1};\,H_{s{=}2}\,]} \;\in\; \mathbb{R}^{32}.
\end{equation}
Descriptors $\{\mathbf{h}\}$ are collected from bit-depths $b{=}7$-$15$ across the training set (scenes and depths), and $K$-means is then applied to obtain the BoE centers $\{\boldsymbol{\mu}_k\}_{k=1}^K$.
With these $K$ centers, the coding-network pool consists of $K{+}1$ models in total, including a base coding network for $b{=}2$-$6$.

\noindent \textbf{BoE Selection Policy:} 
We use a bit-depth-level-dependent policy. For shallow levels ($b\le 6$), we bypass BoE and route to the base coding network. For deeper levels ($b>6$), we select by nearest histogram center:
\begin{equation}
k^{(b)} \;=\; \operatorname*{argmin}_{k\in\{1,\ldots,K\}} \big\|\mathbf{h}^{(b)}-\boldsymbol{\mu}_k\big\|_2,
\end{equation}
where $\mathbf{h}^{(b)}\in\mathbb{R}^{32}$ is the normalized stage-wise occupancy descriptor and $\{\boldsymbol{\mu}_k\}_{k=1}^K$ are the BoE centers.  
The selected index $k^{(b)}$ is signaled in the bitstream so encoder and decoder use the same model at each bit-depth.
By selecting a network according to the bit-depth-specific occupancy distribution, the method reduces the distribution mismatch of single-model designs and attains broad adaptability without assigning a dedicated network to every bit-depth.

\subsection{Training}

Before training, $K$ BoE centers are computed on the training set, and the coding-network pool is instantiated with $K{+}1$ models (one base network plus $K$ BoE models).

Given a point cloud, training proceeds progressively from $b{=}2$ to $b{=}15$.
At each level $b$, the BoE selector chooses one coding network indexed by $k^{(b)} \in \{1,\dots,K{+}1\}$.
The selected network outputs stage-wise occupancy probabilities 
$\{\mathbf{P}_{s,k}^{(b)}\}_{s=1,2}$.
The loss minimizes the total estimated bitrate across stages and levels:
\begin{equation}
\label{eq:train-loss}
\begin{aligned}
\mathcal{L} \;=\; \sum_{b=2}^{15}\sum_{s=1}^{2}\sum_{n=1}^{N_b}
&\; -\log_{2}\mathbf{P}_{s,k}^{(b)}\!\big(n,\, \mathbf{Q}_{s}^{(b)}(n)\big).
\end{aligned}
\end{equation}
Optionally, the loss may be normalized by the input point count to report bits-per-point.

During backpropagation, gradients flow only through the selected network at each level.
Nevertheless, indirect gradients are routed to preceding levels via cross-bit-depth feature propagation, which encourages consistent feature interfaces and stabilizes joint training across the network pool.

%% file: sec/4_experiments.tex
\section{Experimental Results}
\label{sec:exp}
\subsection{Experimental Setups}

\subsubsection{Datasets}
\noindent \textbf{Ford (CTC).}
We adopt the MPEG CTC~\cite{CTC} protocol. The Ford set contains three sequences (01-03), each with 1,500 frames (4,500 total) quantized at 1$mm$ resolution to 18-bit precision. Following RENO~\cite{RENO}, sequence 01 is used for training and sequences 02 and 03 for evaluation.

\noindent \textbf{SemanticKITTI.}
This outdoor LiDAR dataset was captured with a Velodyne HDL-64~\cite{SemanticKITTI}. We quantize coordinates at 1$mm$ resolution, yielding an 18-bit representation. Following~\cite{Unicorn,UniPCGC}, sequences 00-10 are used for training and 11-21 for testing.

For training on both datasets, we voxelize to 4$mm$ to obtain 16 bit-depth point cloud coordinates.

\subsubsection{Implementation Details}
ELiC is implemented in PyTorch~\cite{PyTorch}. 
We use torchac~\cite{torchac} for arithmetic coding and TorchSparse~\cite{torchsppp} for sparse convolutions. 
Training runs 300K iterations with batch size 1.
The learning rate starts at $5\times 10^{-4}$ and is decayed by $0.1$ at 150K and 250K iterations. 
Optimization uses Adam~\cite{Adam}.

We train two variants that differ only in feature channel dimension $D$, namely $D{=}32$ and $D{=}64$, denoted ELiC and ELiC-Large.
All sparse convolutions use $3{\times}3{\times}3$ kernels. 
Unless otherwise specified, BoE uses $K{=}5$ centers; together with the base coding network, the model comprises six coding networks in total.

\subsubsection{Evaluaion and Benchmarks}
We evaluate rate-distortion at five rates using inputs quantized to bit-depths 16, 15, 14, 13, and 12.
This choice follows \cite{martins2024impact}, which shows that in the SemanticKITTI dataset, up to 12-bit precision preserves most points relative to the raw LiDAR data and is sufficient for practical use.

Following the MPEG protocol \cite{schwarz2018emerging}, bitrate is measured in bits per point (BPP) and geometry distortion is reported with point-to-point (D1) and point-to-plane (D2) metrics. 
We also measure encoding/decoding runtime on an RTX 3090 GPU and an Intel Core i9-9900K CPU with 64GB RAM.
All experiments were conducted under WSL2~\cite{barnes2021pro}, which may incur overhead in the reported runtimes.

We compare against state-of-the-art models including RENO and RENO-Large~\cite{RENO}, TopNet~\cite{Wang_2025_CVPR}, Unicorn~\cite{Unicorn}, and G-PCCv30~\cite{GPCC, GPCCv30}.
Following the authors' settings, RENO uses $3{\times}3{\times}3$ kernels with $D{=}32$ channels, while RENO-Large uses $5{\times}5{\times}5$ kernels with $D{=}128$.
For Unicorn, we report results from the official repository. 
For the others, we ran the models and reported our measurements.

Note that all methods losslessly code the input coordinates. 
Consequently, at a fixed input bit-depth, the D1/D2 distortions are identical across methods, and the bitrate is determined solely by the input bit-depth and each method's lossless coding efficiency.

\begin{table*}[!t]
\caption{Per-frame average encoding and decoding time measured on Ford and SemanticKITTI datasets. For TopNet, decoding time is reported over 10 frames due to its extremely long runtime.}
\label{tab:runtime}
\scriptsize
\centering
\begin{tabular}{lcccccccccccccc}
\hline
                                                                                     & \multicolumn{2}{c}{\textbf{G-PCC}}  & \multicolumn{2}{c}{\textbf{RENO}}                                             & \multicolumn{2}{c}{\textbf{RENO-Large}} & \multicolumn{2}{c}{\textbf{Unicorn}} & \multicolumn{2}{c}{\textbf{TopNet}} & \multicolumn{2}{c}{\textbf{ELiC}}                                             & \multicolumn{2}{c}{\textbf{ELiC-Large}} \\
\multirow{-2}{*}{\textbf{\begin{tabular}[c]{@{}l@{}}Input\\ Bit-depth\end{tabular}}} & \textbf{enc (s)} & \textbf{dec (s)} & \textbf{enc (s)}                      & \textbf{dec (s)}                      & \textbf{enc (s)}   & \textbf{dec (s)}   & \textbf{enc (s)}  & \textbf{dec (s)} & \textbf{enc (s)} & \textbf{dec (s)} & \textbf{enc (s)}                      & \textbf{dec (s)}                      & \textbf{enc (s)}   & \textbf{dec (s)}   \\ \hline
\textbf{16 bit}                                                                      & 0.700            & 0.845            & {\color[HTML]{0000FF} 0.173}          & {\color[HTML]{0000FF} 0.179}          & 0.766              & 0.803              & 5.816             & 5.173            & 1.922            & 2357.856         & {\color[HTML]{FF0000} 0.172}          & {\color[HTML]{FF0000} 0.157}          & 0.198              & 0.168              \\
\textbf{15 bit}                                                                      & 0.614            & 0.734            & {\color[HTML]{0000FF} 0.152}          & {\color[HTML]{0000FF} 0.147}          & 0.589              & 0.584              & 4.271             & 3.922            & 1.532            & 2091.421         & {\color[HTML]{FF0000} 0.145}          & {\color[HTML]{FF0000} 0.135}          & 0.162              & 0.143              \\
\textbf{14 bit}                                                                      & 0.540            & 0.600            & {\color[HTML]{0000FF} 0.122}          & {\color[HTML]{0000FF} 0.120}          & 0.434              & 0.423              & 2.842             & 2.516            & 0.917            & 1136.171         & {\color[HTML]{FF0000} 0.121}          & {\color[HTML]{FF0000} 0.112}          & 0.14               & 0.119              \\
\textbf{13 bit}                                                                      & 0.418            & 0.434            & {\color[HTML]{0000FF} \textbf{0.098}} & {\color[HTML]{0000FF} \textbf{0.090}} & 0.299              & 0.290              & 1.899             & 1.767            & 0.608            & 638.485          & {\color[HTML]{FF0000} \textbf{0.095}} & {\color[HTML]{FF0000} \textbf{0.086}} & 0.105              & \textbf{0.096}     \\
\textbf{12 bit}                                                                      & 0.302            & 0.266            & {\color[HTML]{FF0000} \textbf{0.074}} & {\color[HTML]{0000FF} \textbf{0.064}} & 0.207              & 0.197              & 1.661             & 1.473            & 0.322            & 308.066          & {\color[HTML]{FF0000} \textbf{0.074}} & {\color[HTML]{FF0000} \textbf{0.063}} & \textbf{0.084}     & \textbf{0.08}      \\ \hline
\textbf{Avg.}                                                                        & 0.515            & 0.576            & {\color[HTML]{0000FF} 0.124}          & {\color[HTML]{0000FF} 0.120}          & 0.459              & 0.459              & 3.298             & 2.970            & 1.060            & 1306.400         & {\color[HTML]{FF0000} 0.121}          & {\color[HTML]{FF0000} 0.111}          & 0.138              & 0.121              \\ \hline
\textbf{vs. ELiC}                                                                    & 4.24$\times$     & 5.20$\times$     & {\color[HTML]{0000FF} 1.02$\times$}   & {\color[HTML]{0000FF} 1.08$\times$}   & 3.78$\times$       & 4.15$\times$       & 27.16$\times$     & 26.86$\times$    & 8.73$\times$     & 11,811$\times$   & {\color[HTML]{FF0000} 1.00$\times$}   & {\color[HTML]{FF0000} 1.00$\times$}   & 1.14$\times$       & 1.10$\times$       \\ \hline
\end{tabular}
\end{table*}


\subsection{Runtime Efficiency Evaluation}
\label{sec:runtime}
Our primary contribution is achieving high compression efficiency with low computational complexity.
Therefore, we present runtime efficiency first.

Table~\ref{tab:runtime} reports the per-frame average encoding and decoding times on the Ford and SemanticKITTI datasets.
Among all methods, ELiC achieves the fastest performance, with average runtimes of 0.121 seconds for encoding and 0.111 seconds for decoding.
This efficiency stems from an order-preserving coordinate hierarchy that eliminates per-level sorting overhead during up/downscaling and from a streamlined sparse network compared with other methods.

RENO exhibits runtime efficiency similar to ELiC.
In contrast, RENO-Large shows approximately three times higher time complexity than RENO.
While the increased feature dimensionality (32$\rightarrow$128 channels) contributes to this difference, the dominant factor is the use of $5{\times}5{\times}5$ sparse convolutions, which significantly expand the neighborhood search range and memory access volume.
This expansion increases indexing and cache overhead, substantially slowing computation~\cite{chen2023largekernel3d}.

Following ELiC and RENO, ELiC-Large also demonstrates strong runtime efficiency.
Although our models perform bit-depth-wise coding network selection, the selection indices are signaled in the bitstream, so decoding does not incur any additional computational cost.
Consequently, decoding is slightly faster than encoding.
For 12-bit input LiDAR geometry, only RENO, ELiC, and ELiC-Large achieve real-time throughput at 10~FPS, matching the typical LiDAR capture rate.

Unicorn requires over one second per frame for both encoding and decoding.
At each bit-depth level, it employs up to six Inception ResNets~\cite{SparsePCGC} with $5{\times}5{\times}5$ sparse convolutions or k-NN-based neighborhood point attention, which impose heavy computational overhead.


TopNet shows a large encoding-decoding time gap: it encodes within two seconds by using ground-truth occupancy for parallel context extraction, but decoding often exceeds five minutes because it must proceed sequentially at the octree-node level using only prior outputs.

Similarly, G-PCCv30 follows a largely sequential pipeline. 
Despite a pure C++ implementation and a lightweight operation set, the measured runtime exceeded 0.5 seconds for encoding and decoding.

\subsection{Compression Efficiency Evaluation}
\label{sec:compres}

Table~\ref{tab:BDBR} reports BD-Rate~\cite{bjontegaard2001calculation} relative to G-PCCv30~\cite{GPCCv30}, representing the average bitrate savings required to achieve the same quality as G-PCCv30.

On the Ford dataset, ELiC-Large achieves the best compression efficiency, followed by TopNet, Unicorn, ELiC, RENO-Large, and RENO. 
Despite its much lower time complexity, ELiC-Large outperforms all evaluated state-of-the-art models in compression efficiency.
Relative to RENO with comparable runtime efficiency, ELiC achieves more than 8\% higher compression efficiency.
RENO-Large, which increases RENO's feature dimension and kernel size, improves performance by only about 0.7\%.
In contrast, ELiC-Large, which doubles ELiC's feature dimension, provides an additional 3.5\% gain over ELiC.
Considering the runtime results in Table~\ref{tab:runtime}, our models provide a better compression-runtime trade-off than all other methods.

On SemanticKITTI, TopNet attains the best compression efficiency, followed by ELiC-Large, RENO-Large, Unicorn, ELiC, and RENO. 
ELiC-Large delivers compression comparable to TopNet while being far more computationally efficient, yielding a better operating point.
ELiC improves over the RENO by about 8.3\% and attains competitive performance against Unicorn, which operates at more than 20$\times$ higher time complexity. 
Because ELiC has a larger model size (7.75 MB) than RENO (1.10 MB), its performance gains may seem expected.
However, comparing ELiC-Large (30.7 MB) with RENO-Large (78.4 MB) shows the opposite trend: the much smaller ELiC-Large achieves higher compression efficiency and faster runtime, giving it a clear advantage in the size-speed-performance trade-off.

\begin{table}[!t]
\caption{BD-Rate (\%) on Ford and SemanticKITTI relative to the G-PCCv30. D1 = point-to-point, D2 = point-to-plane. Lower is better. Results are based on bit-depths \{16, 15, 14, 13, 12\}.}
\label{tab:BDBR}
\scriptsize
\centering
\begin{tabular}{lcccc}
\hline
                                                                                            & \multicolumn{2}{c}{\textbf{Ford}}                             & \multicolumn{2}{c}{\textbf{KITTI}}                            \\ \cline{2-5} 
\multirow{-2}{*}{\textbf{\begin{tabular}[c]{@{}l@{}}BD-Rate  \\ vs. G-PCCv30\end{tabular}}} & D1 (\%)                       & D2 (\%)                       & D1 (\%)                       & D2 (\%)                       \\ \hline
\textbf{RENO}       & -14.02                        & -14.03                        & -20.90                        & -20.92                        \\
\textbf{RENO-Large} & -14.70                        & -14.70                        & -31.52                        & -31.53                        \\
\textbf{Unicorn}    & -25.41                        & -25.42                        & -29.28                        & -29.30                        \\
\textbf{TopNet}     & {\color[HTML]{0000FF} -26.26} & {\color[HTML]{0000FF} -26.26} & {\color[HTML]{FF0000} -34.10} & {\color[HTML]{FF0000} -34.12} \\
\textbf{ELiC}       & -22.97                        & -22.97                        & -29.23                        & -29.24                        \\
\textbf{ELiC-Large} & {\color[HTML]{FF0000} -26.54} & {\color[HTML]{FF0000} -26.53} & {\color[HTML]{0000FF} -33.26} & {\color[HTML]{0000FF} -33.27} \\ \hline
\end{tabular}
\end{table}

\begin{table}[!t]
\caption{Bitrate savings and relative encoding/decoding runtime ratios of ELiC w/o BoE (1.28 MB) compared to RENO (1.10 MB).}
\label{tab:vsRENO}
\scriptsize
\centering
\begin{tabular}{cccccc}
\hline
\multicolumn{2}{c}{\textbf{Ford}}   & \multicolumn{2}{c}{\textbf{KITTI}}  & \multicolumn{2}{c}{\textbf{Relative Runtime}} \\ \hline
\textbf{D1 (\%)} & \textbf{D2 (\%)} & \textbf{D1 (\%)} & \textbf{D2 (\%)} & \textbf{enc (s)}      & \textbf{dec (s)}      \\ \hline
-8.60            & -8.59            & -6.61            & -6.60            & 0.90$\times$          & 0.85$\times$          \\ \hline
\end{tabular}
\end{table}

\begin{table}[!t]
\caption{Bitrate savings (D1 BD-rates) of ELiC according to BoE pool size $K$, relative to ELiC w/o BoE}
\label{tab:vsWithoutBoE}
\scriptsize
\centering
\begin{tabular}{cccccc}
\hline
\textbf{\begin{tabular}[c]{@{}c@{}}BoE\\ Pool Size\end{tabular}} & \textbf{\begin{tabular}[c]{@{}c@{}}Ford \\ D1 (\%)\end{tabular}} & \textbf{\begin{tabular}[c]{@{}c@{}}KITTI \\ D1 (\%)\end{tabular}} & \textbf{\begin{tabular}[c]{@{}c@{}}BoE\\ Pool Size\end{tabular}} & \textbf{\begin{tabular}[c]{@{}c@{}}Ford \\ D1 (\%)\end{tabular}} & \textbf{\begin{tabular}[c]{@{}c@{}}KITTI \\ D1 (\%)\end{tabular}} \\ \hline
$K=3$                                                            & -1.31                                                            & -2.89                                                             & $K=4$                                                            & -1.55                                                            & -3.06                                                             \\
$K=5$                                                            & {\color[HTML]{0000FF} -1.98}                                     & {\color[HTML]{0000FF} -3.54}                                      & $K=6$                                                            & -1.56                                                            & -3.53                                                             \\
$K=7$                                                            & -1.43                                                            & -3.44                                                             & Per-Level                                                        & {\color[HTML]{FF0000} -2.52}                                     & {\color[HTML]{FF0000} -4.32}                                      \\ \hline
\end{tabular}
\end{table}

\begin{figure*}[!t]
\centering

\begin{minipage}[b]{0.05\textwidth}\centering
  \rotatebox{90}{\footnotesize \qquad\textbf{15 bit-depth level}}
\end{minipage}\hspace{0.01\textwidth}
\begin{subfigure}[b]{0.28\textwidth}
  \includegraphics[width=\textwidth]{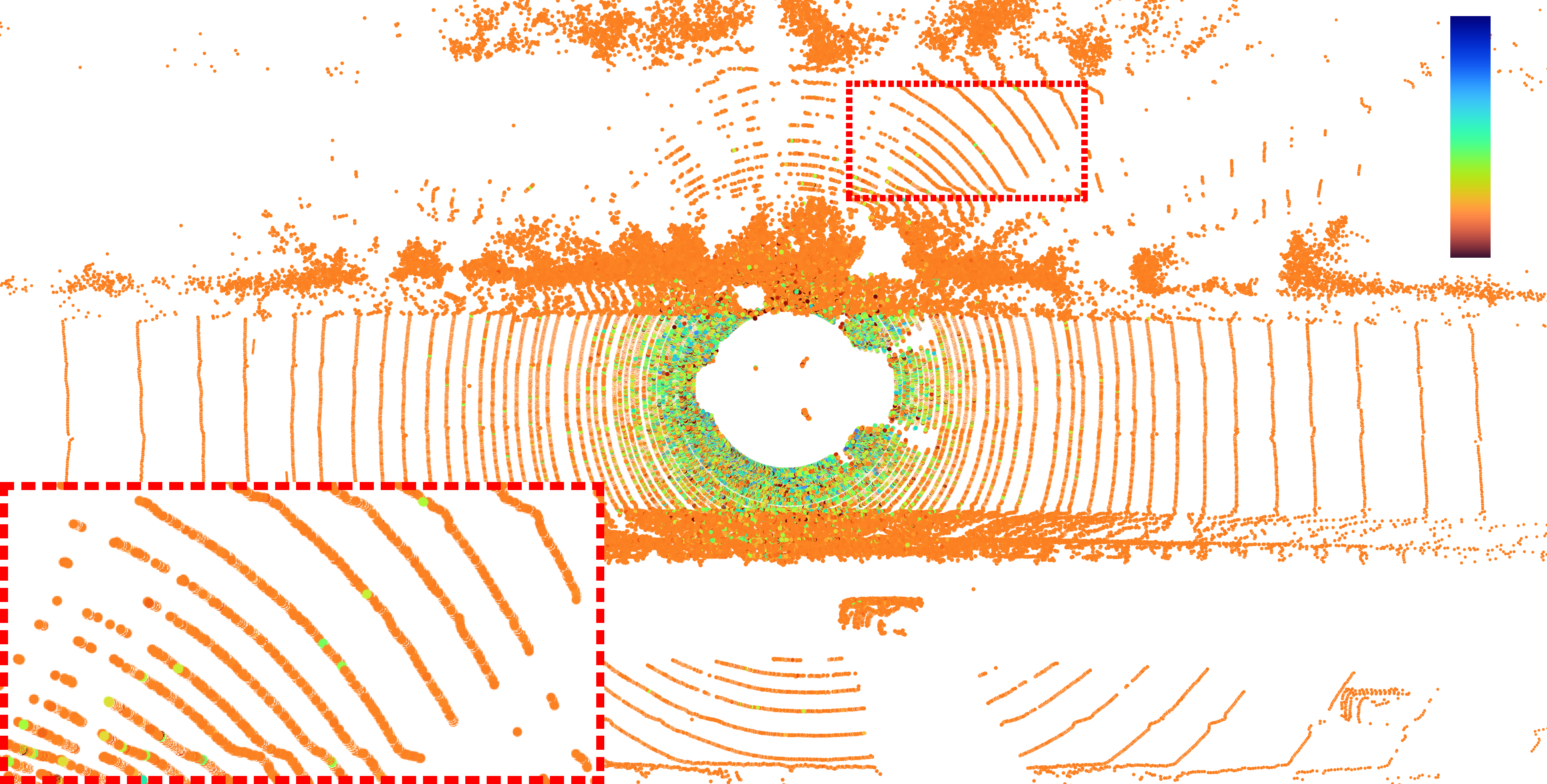}
  \caption{RENO, 353 KBits}
\end{subfigure}\hspace{0.01\textwidth}
\begin{subfigure}[b]{0.28\textwidth}
  \includegraphics[width=\textwidth]{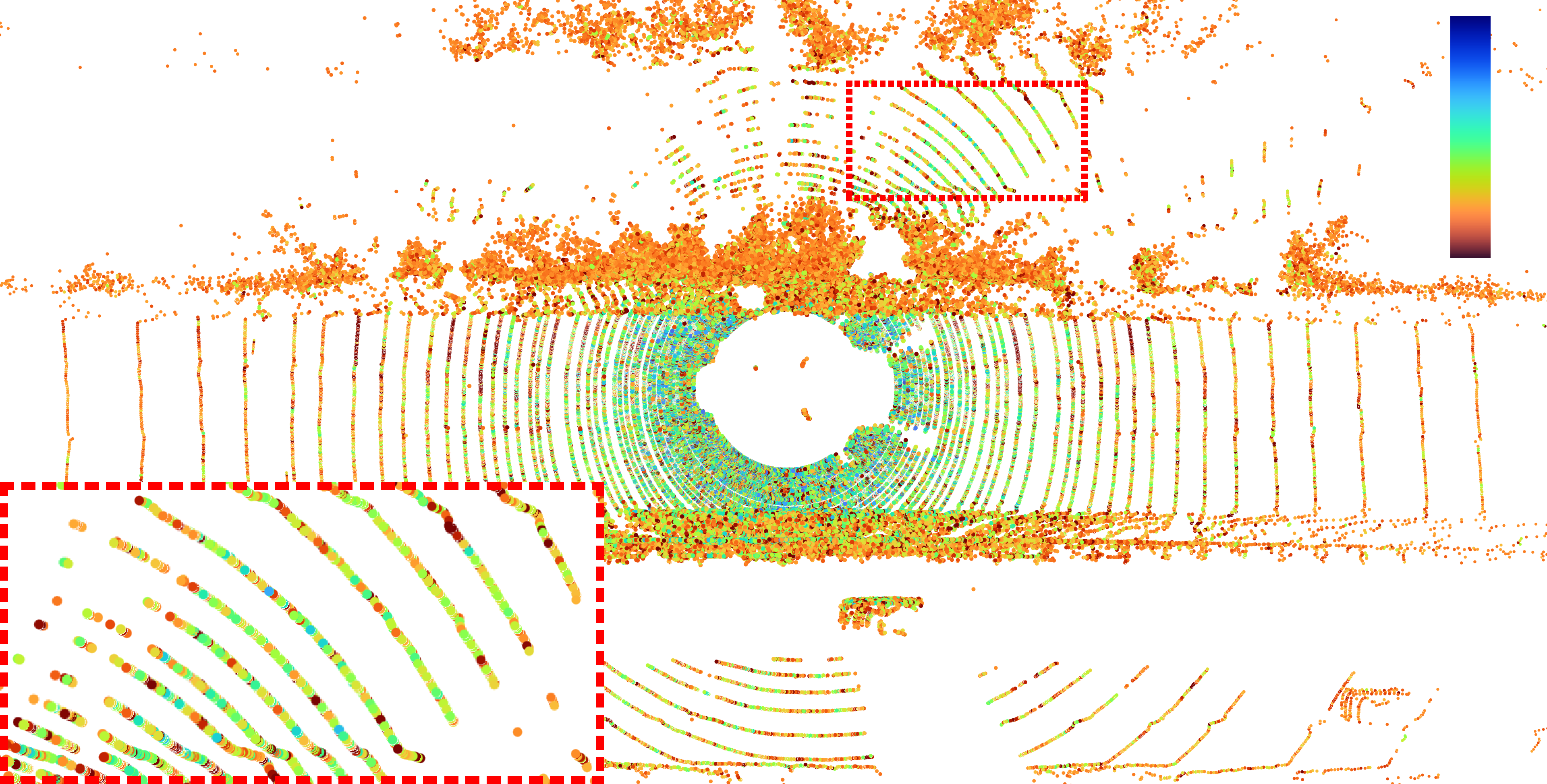}
  \caption{ELiC w/o BoE, 322 KBits}
\end{subfigure}\hspace{0.01\textwidth}
\begin{subfigure}[b]{0.28\textwidth}
  \includegraphics[width=\textwidth]{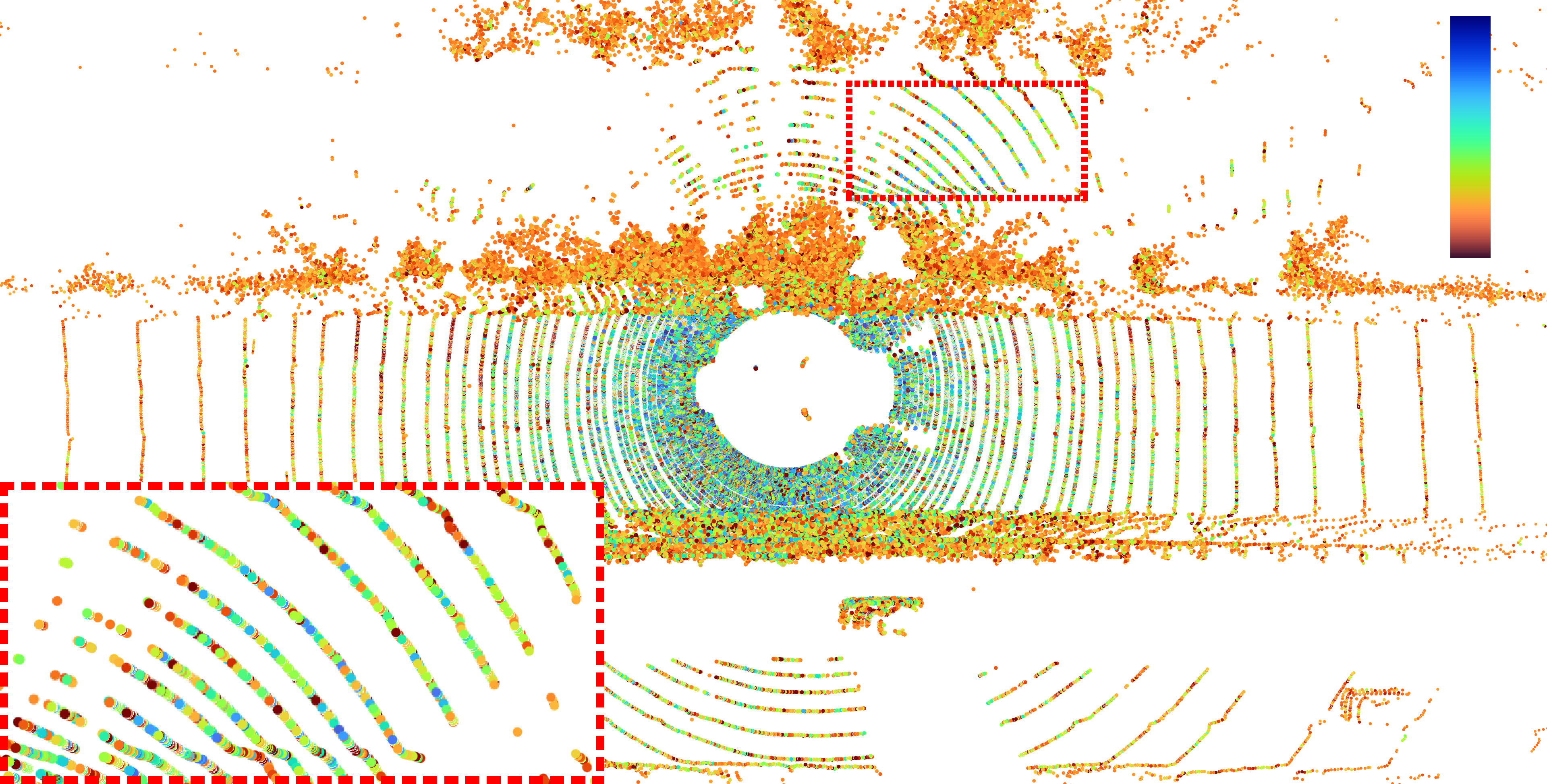}
  \caption{ELiC, 308 KBits}
\end{subfigure}

\vspace{0.015\textwidth}

\begin{minipage}[b]{0.05\textwidth}\centering
  \rotatebox{90}{\footnotesize \qquad\textbf{12 bit-depth level}}
\end{minipage}\hspace{0.01\textwidth}
\begin{subfigure}[b]{0.28\textwidth}
  \includegraphics[width=\textwidth]{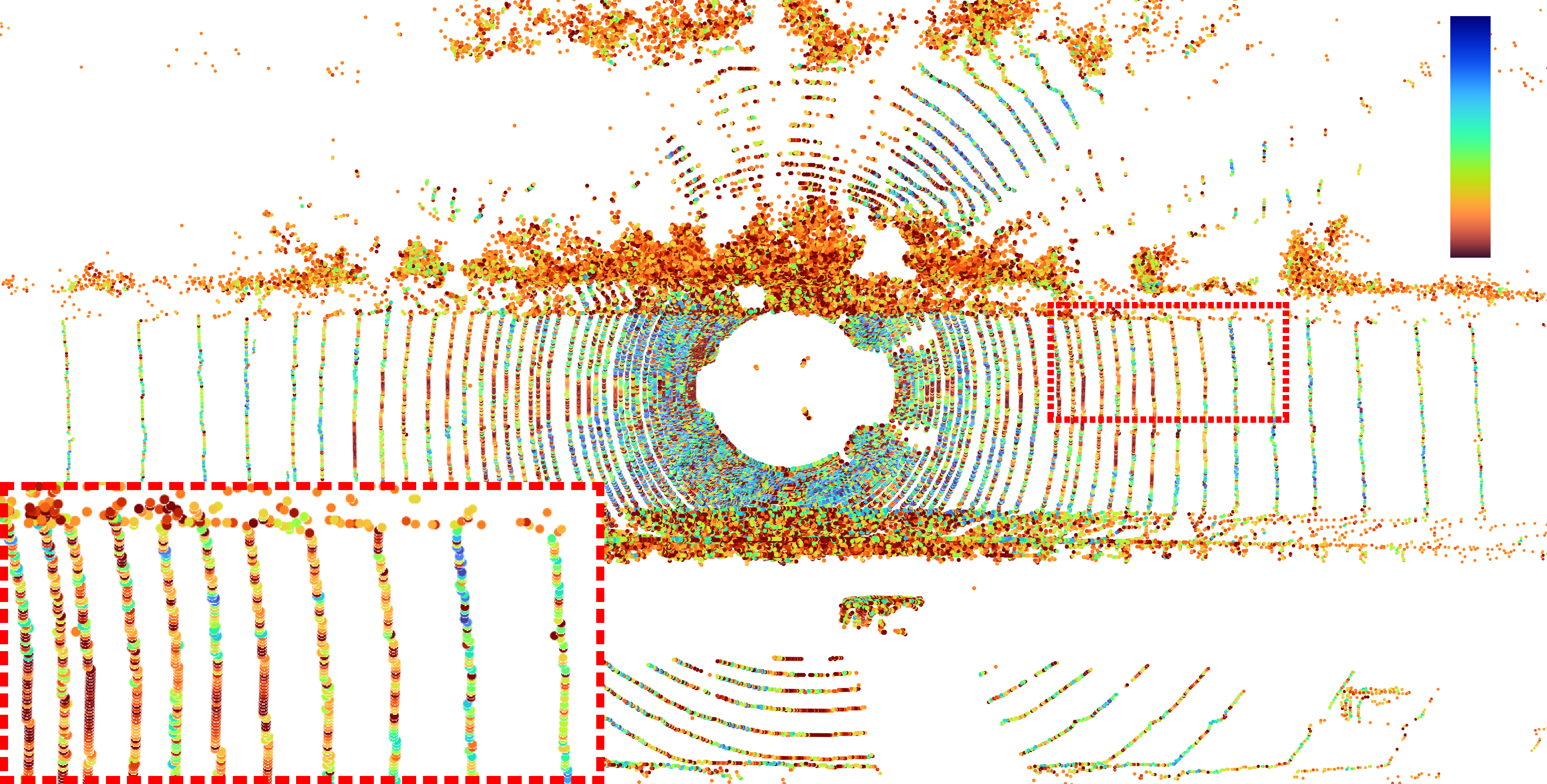}
  \caption{RENO, 253 KBits}
\end{subfigure}\hspace{0.01\textwidth}
\begin{subfigure}[b]{0.28\textwidth}
  \includegraphics[width=\textwidth]{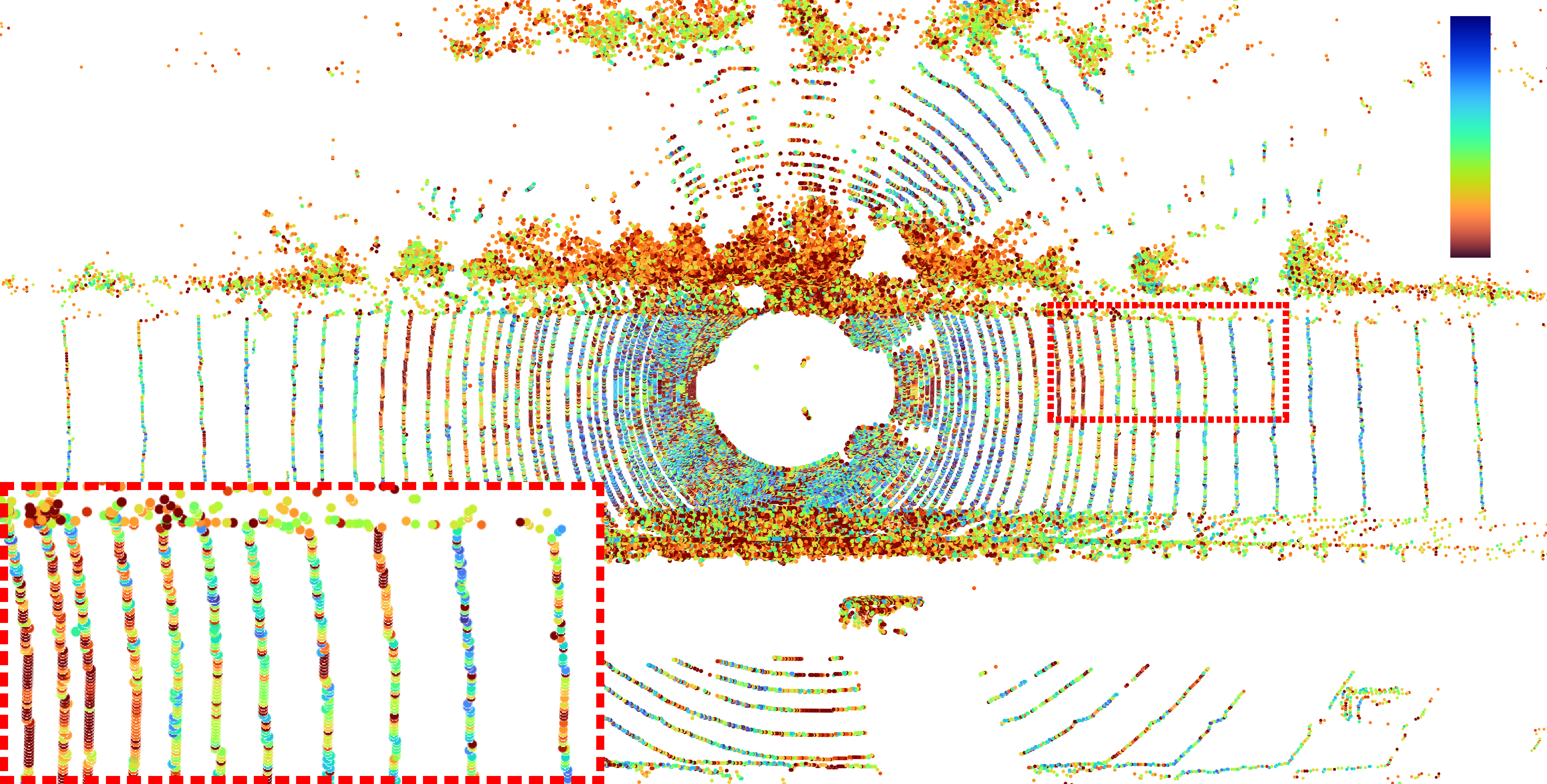}
  \caption{ELiC w/o BoE, 238 KBits}
\end{subfigure}\hspace{0.01\textwidth}
\begin{subfigure}[b]{0.28\textwidth}
  \includegraphics[width=\textwidth]{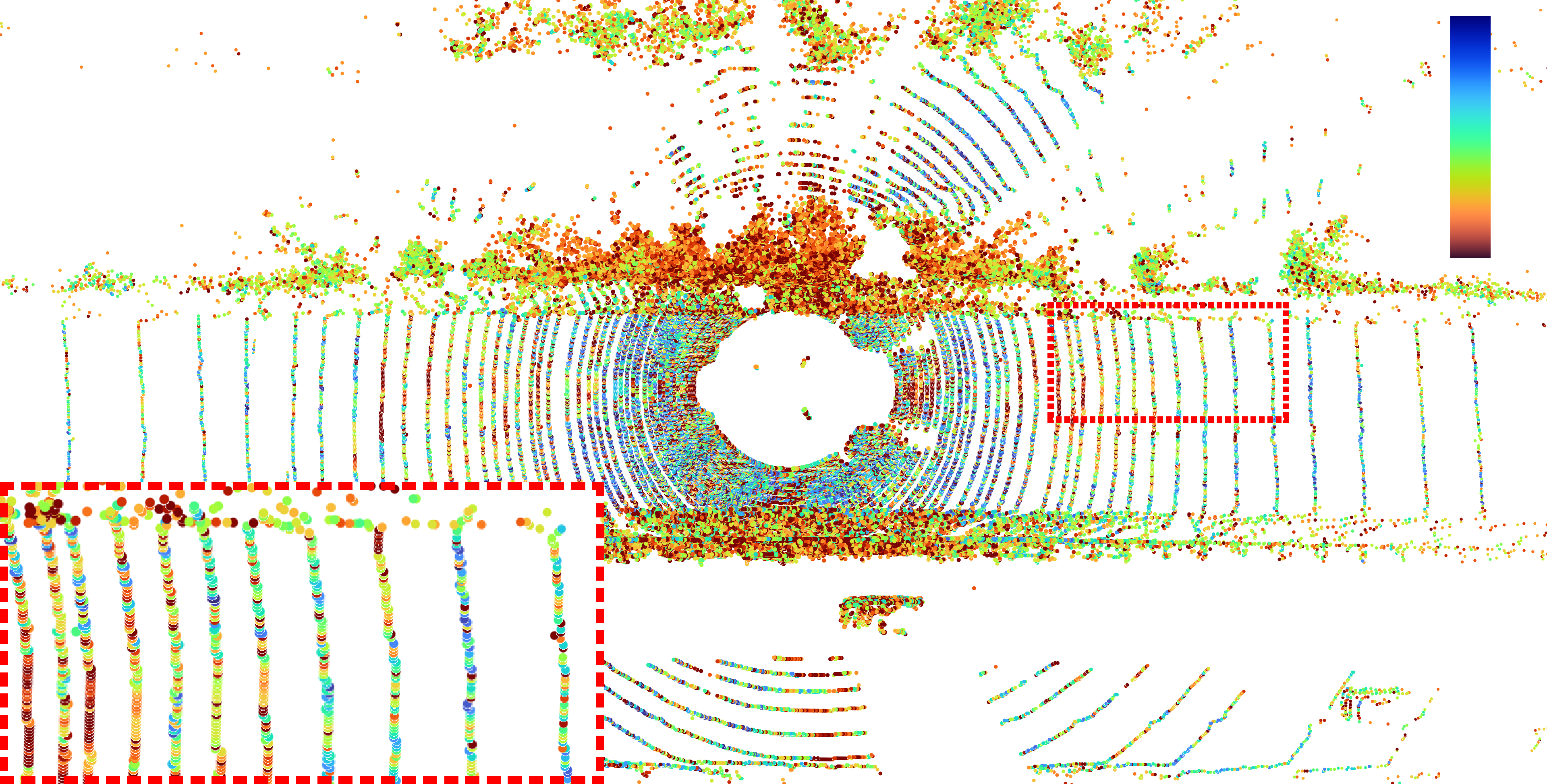}
  \caption{ELiC, 229 KBits}
\end{subfigure}

\caption{Per-point bit allocation on SemanticKITTI frame at the 15 and 12 bit-depth levels for RENO, ELiC w/o BoE, and ELiC ($K{=}5$). The colormap encodes lower (blue) to higher (dark orange) predicted bits per point.} 
\label{fig:bitalloc}

\end{figure*}

\subsection{Ablation Studies}
\noindent \textbf{ELiC without BoE.} To isolate the effect of cross-bit-depth feature propagation, we evaluate ELiC with a single coding network and no BoE selection.
Table~\ref{tab:vsRENO} reports bitrate savings and relative runtime versus RENO.
Encoding is about 10\% faster and decoding about 15\% faster than RENO, while ELiC without BoE achieves 8.6\% bitrate savings on Ford and 6.6\% on SemanticKITTI.
The gain arises from the change in information flow: RENO re-derives context independently at each level, whereas ELiC propagates lower-level context upward.
Fig.~\ref{fig:bitalloc} provides supporting evidence.
At the 15 bit-depth level, the per-point bit allocation of RENO fails to reduce predicted bit usage efficiently outside the dense sensor-center region.
In contrast, ELiC without BoE lowers the predicted bit usage even in sparse peripheral regions, indicating that context information accumulated from lower bit-depths is effectively propagated upward and utilized for encoding at highly sparse levels.
Thus, even without BoE, cross-bit-depth feature propagation alone yields substantial compression improvements with minimal runtime overhead.

\noindent \textbf{Effect of BoE Pool Size.} We evaluate how performance scales with the number of coding networks in the BoE pool.
Excluding the base coding network, we vary the pool size $K\in\{3,4,5,6,7\}$ and report BD-Rate relative to ELiC without BoE.
As an upper-bound reference, we also test a per-level variant with 14 dedicated coding networks (one for each bit-depth level from $b{=}2$ to $b{=}15$).

Table~\ref{tab:vsWithoutBoE} summarizes the results. 
Assigning a dedicated coding network to each bit-depth yields the best compression. 
Even a small pool helps. With BoE at $K{=}3$, bitrate savings over ELiC without BoE are 1.31\% on Ford and 2.89\% on SemanticKITTI. 
Performance increases up to $K{=}5$ and then slightly drops.
Compared with the upper-bound model with 14 dedicated networks, BoE with $K{=}5$ is within 0.66\% BD-Rate loss while using only $6/14$ of the network capacity, delivering stronger compression performance per model size.

\noindent \textbf{Runtime Impact of Morton-Order Hierarchy.} Table~\ref{tab:morton_runtime1} compares the runtime of ELiC with and without the Morton-order--preserving hierarchy.
Switching from Morton-order traversal to per-level explicit sorting, as in prior methods, increases average encoding and decoding latency by about 14.8\% and 13.3\%, respectively. 
This confirms that maintaining a hierarchical Morton order removes sorting overhead and improves overall runtime efficiency.

\begin{table}[t!]
\caption{Runtime comparison between ELiC with explicit sorting and ELiC with Morton-order-preserving hierarchy.}
\label{tab:morton_runtime1}
\centering
\scriptsize
\begin{tabular}{cccccc}
\hline
\multicolumn{2}{c}{\textbf{\begin{tabular}[c]{@{}c@{}}ELiC with\\ Explicit Sorting\end{tabular}}} & \multicolumn{2}{c}{\textbf{\begin{tabular}[c]{@{}c@{}}ELiC with \\ Morton-Order\end{tabular}}} & \multicolumn{2}{c}{\textbf{\begin{tabular}[c]{@{}c@{}}Latency\\ Reduction\end{tabular}}} \\
\textbf{enc (s)}                                & \textbf{dec (s)}                                & \textbf{enc (s)}                               & \textbf{dec (s)}                              & \textbf{enc (s)}                            & \textbf{dec (s)}                           \\ \hline
0.142                                           & 0.128                                           & 0.121                                          & 0.111                                         & -0.021                                      & -0.017                                     \\ \hline
\end{tabular}
\end{table}

%% file: sec/5_conclusion.tex
\section{Limitations and Conclusion}
\label{sec:limit_conclusion}

We propose ELiC, which combines Morton-order-preserving hierarchy, cross–bit-depth feature propagation, and adaptive BoE network selection to improve LiDAR geometry compression while retaining real-time efficiency. 
The method delivers strong compression performance on Ford and SemanticKITTI with low runtime. 

However, all BoE models currently rely on a fixed 4+4 two-stage factorization, which can be suboptimal in denser regimes where bit grouping misaligns with spatial correlation. 
Addressing this likely requires an input-distribution–adaptive factorization or a joint BoE that selects both the network and its bit-split strategy.

Looking forward, we will pursue quantization-aware training~\cite{rokh2023comprehensive} and knowledge distillation~\cite{chen2021distilling} to reduce compute and power, hardware-specific inference optimizations~\cite{jeong2022tensorrt} for deployment, and extensions to streaming, multi-sensor, and task-aware pipelines that jointly optimize compression with downstream perception.

%% file: sec/X_suppl.tex
\clearpage
\setcounter{page}{1}

\renewcommand{\thesection}{\Alph{section}}
\setcounter{section}{0}

\setcounter{figure}{0}
\renewcommand{\thefigure}{S\arabic{figure}}

\setcounter{table}{0}
\renewcommand{\thetable}{S\arabic{table}} 

\maketitlesupplementary

\section{Encoding and Decoding Pipelines}

\begin{figure}[!th]
  \centering
  \begin{subfigure}[t]{\linewidth}
    \centering
    \includegraphics[width=\linewidth, trim=40 0 40 0, clip]{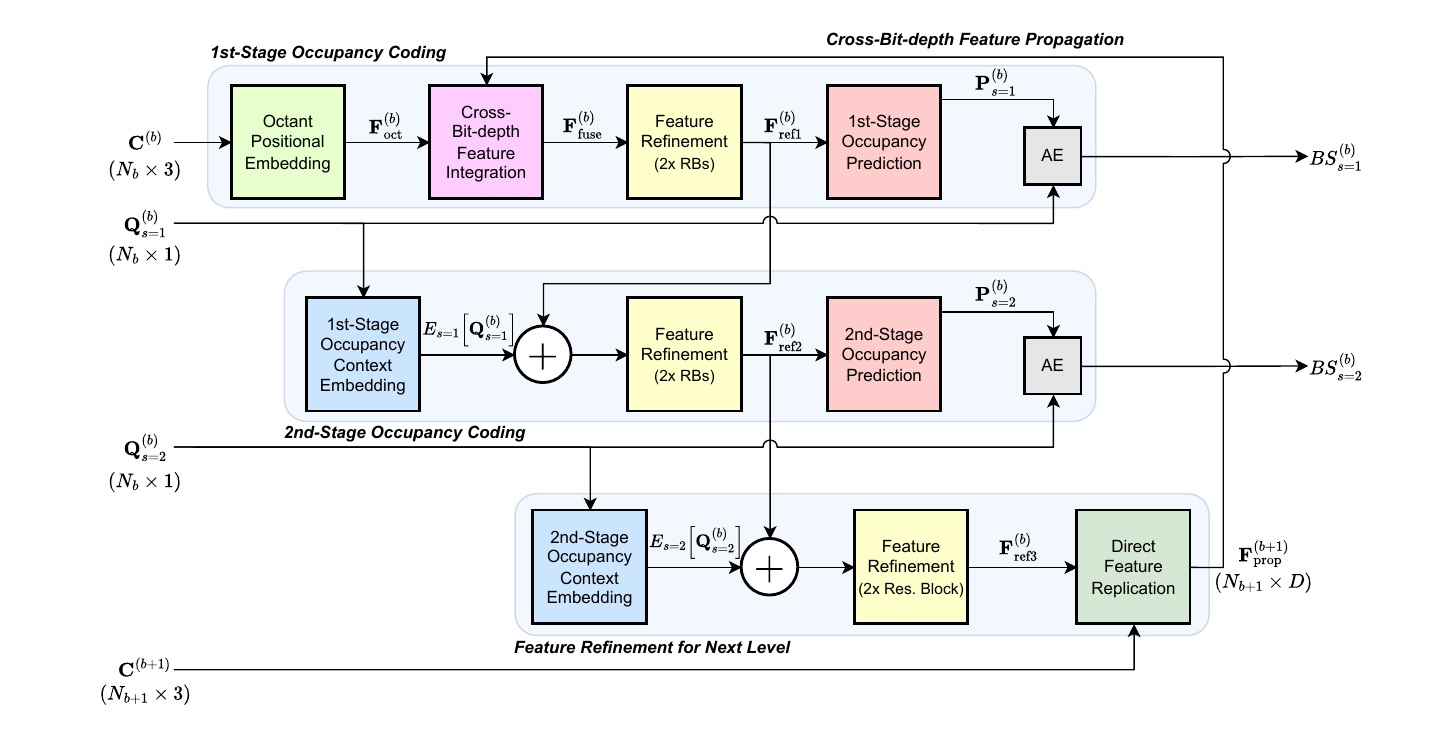}
    \caption{Encoding pipeline}
  \end{subfigure}
  
  \begin{subfigure}[t]{\linewidth}
    \centering
    \includegraphics[width=\linewidth, trim=40 0 40 0, clip]{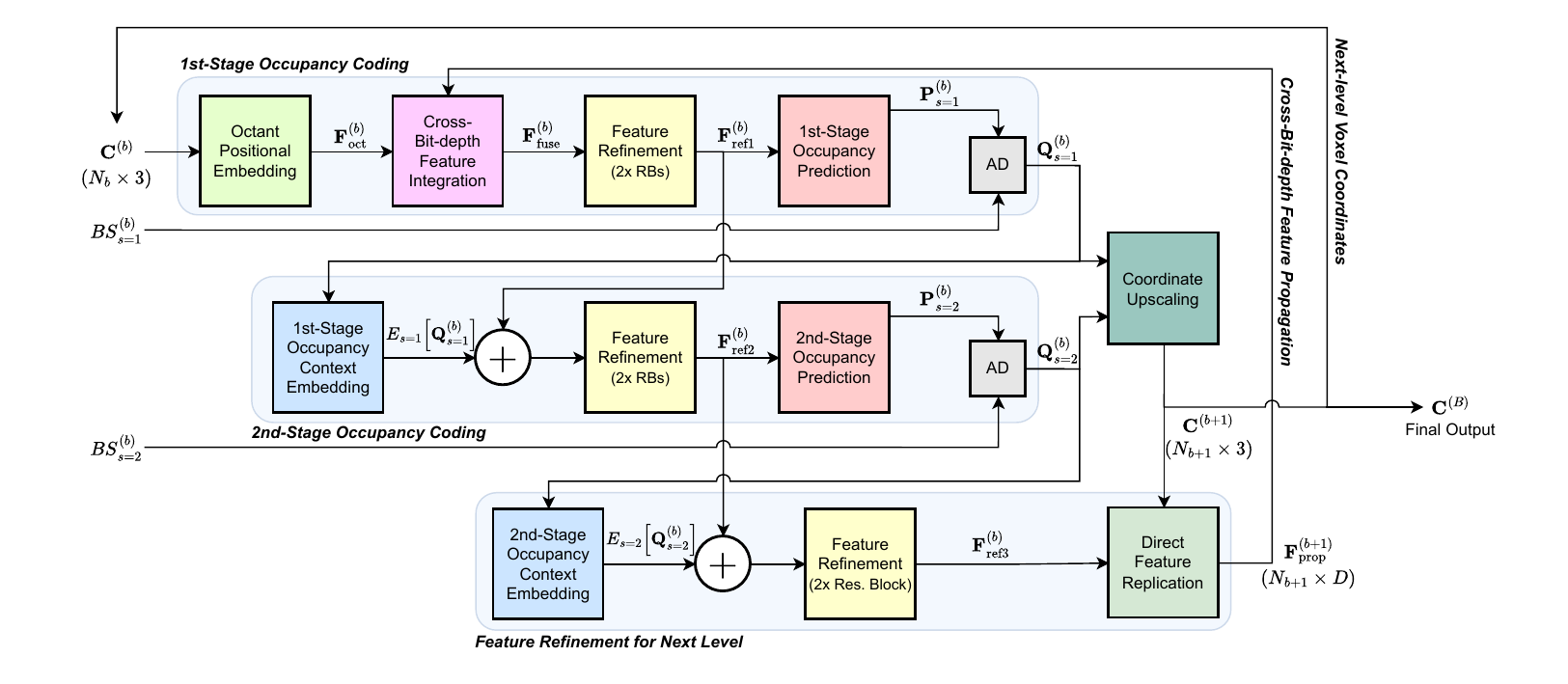}
    \caption{Decoding pipeline}
  \end{subfigure}
  \caption{Coding network execution pipeline for ELiC during encoding and decoding.}
  \label{fig:pipeline}
\end{figure}

Fig.~\ref{fig:pipeline} illustrates the ELiC coding-network execution pipeline for encoding and decoding.
Below we describe the encoding and decoding flows.

At each bit-depth level $b$, the encoder receives the voxel coordinates $\mathbf{C}^{(b)}$ and the stage-wise quadrant labels $\{\mathbf{Q}_{s}^{(b)}\}_{s{=}1,2}$.
Following the selection policy in Sec.~\ref{sec:BoE}, the index $k^{(b)}$ of the coding network to use at bit-depth level $b$ is determined.
The selected network processes the inputs and produces the estimated occupancy probabilities
$\{\mathbf{P}_{\text{s}}^{(b)}\}_{\text{s}=1,2}$.
These probabilities are then passed to the arithmetic encoder, which generates the bitstreams $\{BS_{s}^{(b)}\}_{s{=}1,2}$.

This procedure is repeated sequentially for $b{=}2$ to $B{-}1$.
The final compressed representation consists of the base coordinates $\mathbf{C}^{(2)}$, the set of stage-wise bitstreams $\{BS_{s}^{(b)}\}_{s{=}1,2}$, and the BoE indices $\{k^{(b)}\}_{b=7}^{B-1}$ (for $b{=}2,{\dots},6$ the base coding network is used).

At decoding time, the process mirrors the encoder.
The decoder is given the base coordinates $\mathbf{C}^{(2)}$ and initializes $\mathbf{C}^{(b)}$ for $b{=}2$.

For each bit-depth level $b{=}2$ to $B{-}1$, the decoder first reads the BoE index $k^{(b)}$ from the bitstream, selecting the same coding network as the encoder.
Given $\mathbf{C}^{(b)}$, the selected network predicts the stage-wise quadrant probability distributions $\{\mathbf{P}_{s}^{(b)}\}_{s{=}1,2}$.
These probabilities are then used by the arithmetic decoder to reconstruct the quadrant labels $\{\mathbf{Q}_{s}^{(b)}\}_{s{=}1,2}$ from the bitstreams $\{BS_{s}^{(b)}\}_{s{=}1,2}$.

Once both stages are decoded, the voxel coordinates are expanded to the next bit-depth, and $\mathbf{C}^{(b+1)}$ is generated from the child-occupancy map $\mathbf{O}^{(b)}{=}16\,\mathbf{Q}_{s=2}^{(b)}{+}\mathbf{Q}_{s=1}^{(b)}$.
This procedure repeats until $b{=}B{-}1$, at which point the final reconstructed point set $\mathbf{C}^{(B)}$ is obtained.

\section{Comparison of Cross-Bit-depth and Level-Independent Architectures}

\begin{figure}[h!]
  \centering
  \begin{subfigure}{\linewidth}
    \centering
    \includegraphics[width=\linewidth, trim=10 0 10 0, clip]{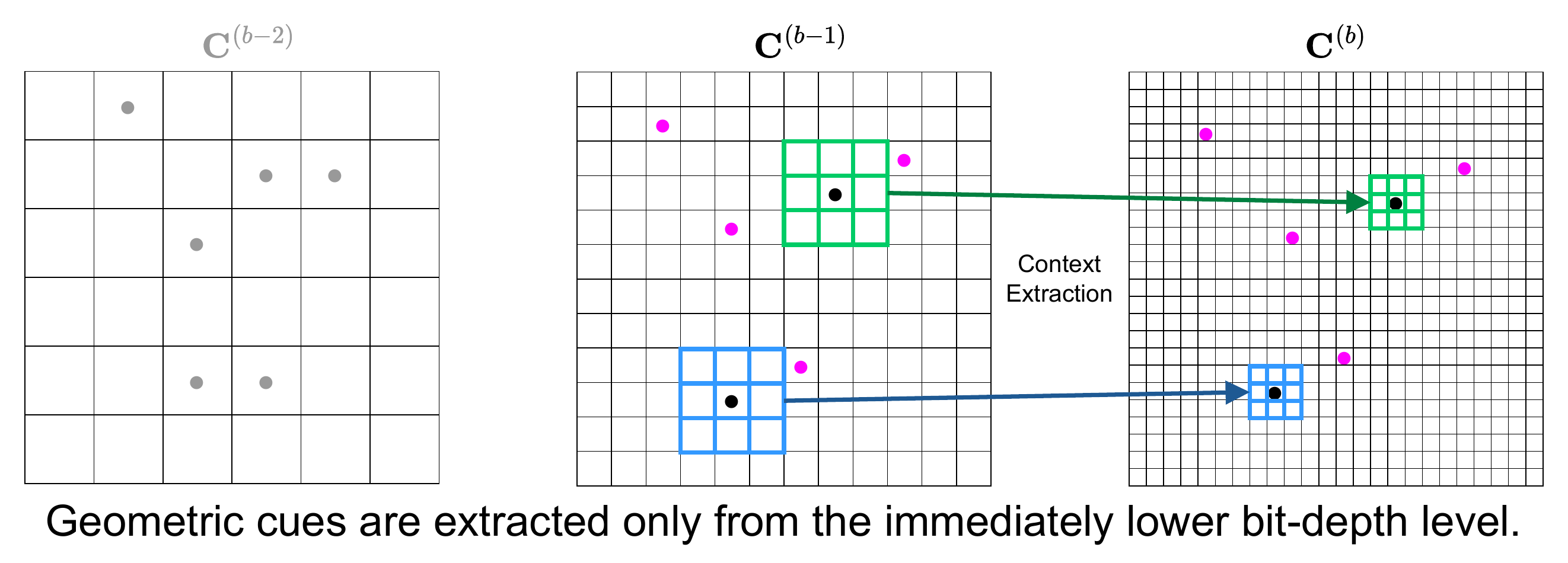}
    \caption{Bit-depth-level-independent design}
  \end{subfigure}
  
  \begin{subfigure}{\linewidth}
    \centering
    \includegraphics[width=\linewidth, trim=10 0 10 0, clip]{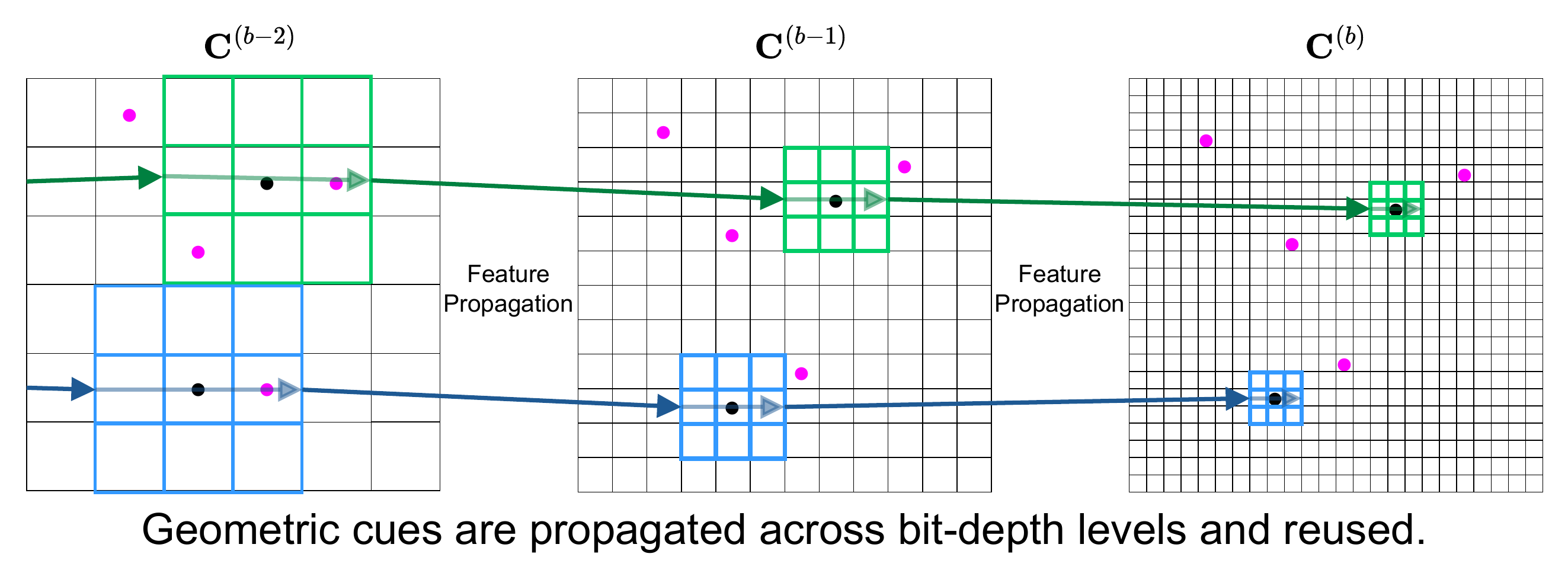}
    \caption{Cross-bit-depth feature propagation design}
  \end{subfigure}
  \caption{Comparison between level-independent and cross-bit-depth designs for LiDAR geometry compression. 
  (a) Bit-depth-level-independent design: each level re-derives lower-level context solely from the current-level voxel coordinates.
  (b) Cross-bit-depth feature propagation: features extracted at lower, denser levels are propagated and fused at higher levels.}
  \label{fig:cross}
\end{figure}

Fig.~\ref{fig:cross} compares the bit-depth-level-independent design with ELiC’s cross-bit-depth feature propagation design.

In the bit-depth-level-independent design, each bit-depth level computes $\mathbf{C}^{(b-1)}$ or $\mathbf{O}^{(b-1)}$  from the current voxel coordinates $\mathbf{C}^{(b)}$ and re-derives lower-level occupancy context from them.
A drawback is that when both the current and lower levels are highly sparse, the available occupancy context carries minimal predictive information. 

In contrast, the cross-bit-depth feature propagation design propagates features from a lower level to the next level.
The propagated features are blended with the current features via Eq.~\ref{eq:cs-fuse} of the main paper and then passed upward again, with the blending ratios determined by channel-wise weights learned during training.
By naturally propagating features from coarse to fine resolutions, the model preserves informative multi-scale structure even when $\mathbf{C}^{(b)}$ and $\mathbf{C}^{(b-1)}$ are extremely sparse, thereby strengthening occupancy prediction and improving entropy-modeling robustness.

\section{Morton-Order-Preserving Coordinate Transformation}
\label{sec:zorder1}

By enforcing a Morton-order-preserving hierarchy, we eliminate the expensive up- and down-convolution operations across the entire encoding and decoding pipeline that are typically used to construct the bit-depth hierarchy and its occupancy labels.
Our method requires only a single Morton-order sort of the input voxel coordinates before encoding, after which all subsequent coordinate and occupancy labels are obtained via cheap integer operations.

Let $\mathbf{C}^{(b)}=\{\mathbf{c}_i^{(b)}=(x_i^{(b)},y_i^{(b)},z_i^{(b)})\}_{i=1}^{N_b}$ denote the occupied voxel coordinates at bit-depth $b$.  
Each voxel is assigned a 3D Morton code
\begin{equation}
m(\mathbf{c}^{(b)})=\sum_{k=0}^{b-1}\big(x_k^{(b)}+(y_k^{(b)}\!\ll\!1)+(z_k^{(b)}\!\ll\!2)\big)8^k,
\end{equation}
where $(x_k^{(b)},y_k^{(b)},z_k^{(b)})$ are the $k$-th least significant bits of each coordinate.  
Sorting by $m(\mathbf{c})$ yields a Z-ordered sequence in which spatial adjacency is preserved along a one-dimensional index.

A duplicated set of parent coordinates is obtained by halving the integer coordinates,
\begin{equation}
\widetilde{\mathbf{C}}^{(b-1)} = \Big\lfloor \frac{\mathbf{C}^{(b)}}{2} \Big\rfloor,
\end{equation}
which is equivalent to truncating the three least significant bits of the Morton code:
\begin{equation}
m^{(b-1)}(\widetilde{\mathbf{c}}^{(b-1)}) = m^{(b)}(\mathbf{c}^{(b)}) \gg 3.
\end{equation}
Therefore, when $\mathbf{C}^{(b)}$ is sorted in Morton order, removing duplicates from each consecutive group of identical entries in $\widetilde{\mathbf{C}}^{(b-1)}$ directly yields the unique parent coordinates $\mathbf{C}^{(b-1)}$.

Because the three-bit right shift preserves lexicographic order on Morton codes, the parent coordinates inherit the same Morton ranking as their children. 
As a result, all voxels that share an identical parent code form a contiguous interval in this order. 
For each parent voxel, we then define an occupancy label
\begin{equation}
\begin{aligned}
\mathbf{O}^{(b-1)}_j &= \sum_{\mathbf{c}_i^{(b)}\in\mathrm{child}(\mathbf{c}_j^{(b-1)})} 2^{\,\pi(\mathbf{c}_i^{(b)})},\\
\pi(\mathbf{c}_i^{(b)}) &= (x_i^{(b)}\bmod2)+2(y_i^{(b)}\bmod2)+4(z_i^{(b)}\bmod2),
\end{aligned}
\end{equation}
where $\pi(\mathbf{c}_i^{(b)})\in\{0,\ldots,7\}$ assigns each child to one of the eight local octants in the $2{\times}2{\times}2$ voxel block centered at $\mathbf{c}_j^{(b-1)}$. The resulting 8-bit integer $\mathbf{O}^{(b-1)}_j$ therefore encodes exactly which of these eight child positions are occupied.

In practice, this occupancy-label generation uses only inexpensive integer operations on the child coordinates and avoids the conventional \(2{\times}2{\times}2\) down-convolution with fixed kernel weights \([1, 2, 4, 8, 16, 32, 64, 128]\) and the associated kernel-map construction overhead.

Conversely, the finer level can be reconstructed directly from $(\mathbf{C}^{(b-1)},\mathbf{O}^{(b-1)})$ by expanding every parent voxel according to its occupancy bits.
Let the fixed octant offsets be $\{\boldsymbol{\delta}_u\}_{u=0}^{7}\subset\{0,1\}^3$ listed in Morton octant order
$u=\delta_x+2\delta_y+4\delta_z$.
Given parent coordinates $\mathbf{C}^{(b)}=\{\mathbf{c}^{(b)}_n\}_{n=1}^{N_b}$ and 8-bit occupancies $\mathbf{O}^{(b)}\in\{0,\ldots,255\}^{N_b}$, form the candidate set
\begin{equation}
\widetilde{\mathbf{C}}^{(b+1)} \;=\; 
\big\{\, 2\,\mathbf{c}^{(b)}_n + \boldsymbol{\delta}_u \;\big|\; n\in[1\!:\!N_b],\, u\in[0\!:\!7] \,\big\},
\end{equation}
i.e., for each parent, enumerate its eight child candidates in a fixed (Morton) octant order.
Define a Boolean mask
\begin{equation}
\label{eq:maskgen}
\mathbf{M}^{(b)}(n,u)\;=\;\big((\mathbf{O}^{(b)}_n \gg u)\,\&\,1\big)\in\{0,1\},
\end{equation}
which selects the octant $u$ when its occupancy bit is set. 
Since the mapping from each occupancy value in $\{0,\ldots,255\}$ to its 8-bit pattern is fixed, $\mathbf{M}^{(b)}$ can be implemented more efficiently via a precomputed lookup table, avoiding repeated bit-shift and mask operations.
Applying this mask by row selection yields
\begin{equation}
\mathbf{C}^{(b+1)} \;=\; \big\{\, \widetilde{\mathbf{C}}^{(b+1)}(n,u) \; \big|\; \mathbf{M}^{(b)}(n,u)=1 \,\big\}.
\end{equation}
Equivalently, the same selection can be expressed in vectorized form.  
Flattening $\mathbf{M}^{(b)}$ in row-major order yields a vector in $\{0,1\}^{8N_b}$, which we use to index $\widetilde{\mathbf{C}}^{(b+1)}$ and obtain the actual child coordinates:
\begin{equation}
\mathbf{C}^{(b+1)}
= \widetilde{\mathbf{C}}^{(b+1)}\big|_{\mathbf{M}^{(b)}=1}.
\end{equation}
If per-parent features $\mathbf{F}^{(b)}\in\mathbb{R}^{N_b\times D}$ are propagated, replicate them to the candidate set
and select with the same mask:
\begin{equation}
\begin{aligned}
\widetilde{\mathbf{F}}^{(b)}(n,u) &= \mathbf{F}^{(b)}(n),\\
\mathbf{F}^{(b+1)}_{\text{prop}} &= \big\{\,\widetilde{\mathbf{F}}^{(b)}(n,u)\;\big|\;\mathbf{M}^{(b)}(n,u)=1\,\big\},
\end{aligned}
\end{equation}
which can equivalently be written in vectorized form as
\begin{equation}
\mathbf{F}_{\text{prop}}^{(b+1)} \;=\; \widetilde{\mathbf{F}}^{(b)}|_{\mathbf{M}^{(b)}=1}\;\in\mathbb{R}^{N_{b+1}\times D}.
\end{equation}
Since parents are listed in Morton order and $\{\boldsymbol{\delta}_u\}_{u=0}^{7}$ is enumerated in Morton octant order,
the selection by $\mathbf{M}^{(b)}$ preserves the global Morton order for both $\mathbf{C}^{(b+1)}$ and
$\mathbf{F}^{(b+1)}_{\text{prop}}$ without any re-sorting.

The reference implementation of this Morton-order-preserving hierarchy and occupancy labeling procedure is available in the \texttt{morton.py} file in the code repository.

\section{Runtime Impact of Morton-Order Hierarchy}
\label{sec:zorder2}

\begin{figure}[h!]
  \centering
  \begin{subfigure}{\linewidth}
    \centering
    \includegraphics[width=\linewidth, trim=80 0 80 0, clip]{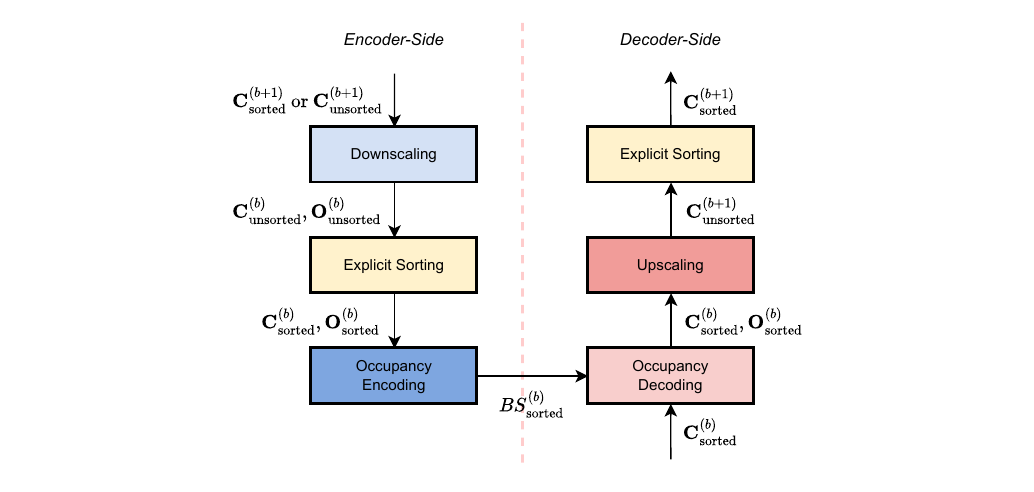}
    \caption{With explicit per-level sorting}
  \end{subfigure}
  
  \begin{subfigure}{\linewidth}
    \centering
    \includegraphics[width=\linewidth, trim=80 0 80 0, clip]{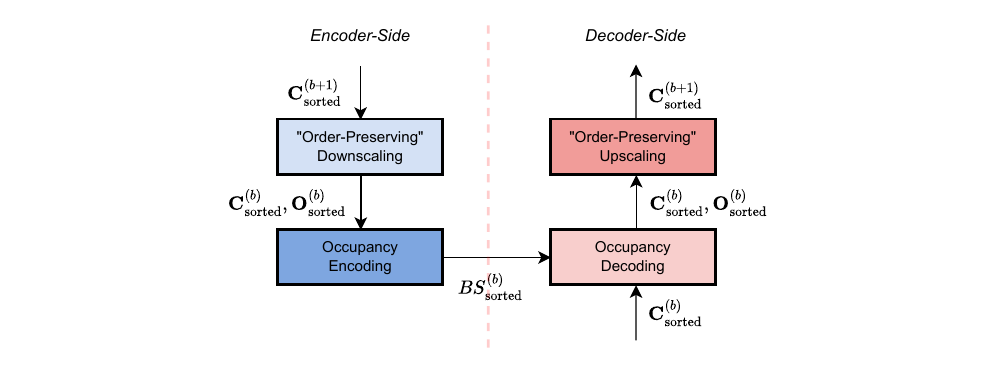}
    \caption{With Morton-order-preserving hierarchy}
  \end{subfigure}
  \caption{Comparison between explicit per-level sorting and a Morton-order-preserving hierarchical representation. 
  (a) Explicit sorting reorders points at each bit-depth level, incurring additional overhead in both encoding and decoding. 
  (b) A Morton-order-preserving hierarchy maintains consistent spatial ordering across levels, enabling stable parent-child mapping and reducing per-level processing cost.}
  \label{fig:ordering}
\end{figure}

Fig.~\ref{fig:ordering} provides a high-level comparison, at a single bit-depth level during compression, between explicit per-level sorting and our Morton-order-preserving hierarchy.
In sparse-tensor-based geometry compression, it is essential that the encoder and decoder operate on exactly the same coordinate order so that occupancy bits are generated and consumed with aligned voxel indices.
Existing sparse-tensor-based approaches satisfy this by performing identical sorting at every downscaling and upsampling step in both encoding and decoding. 
In contrast, ELiC performs a single initial 3D Morton sort of the input voxel coordinates $\mathbf{C}^{(B)}$ and then applies downscaling and upsampling that preserve this global order, eliminating the per-level re-sorting required by prior methods in both encoding and decoding.

Table~\ref{tab:morton_runtime2} compares the runtime of ELiC with and without the Morton-order-preserving hierarchy.
When explicit sorting is replaced by Morton-order traversal, the average encoding and decoding times are reduced from 0.142 seconds and 0.128 seconds to 0.121 seconds and 0.111 seconds, respectively—corresponding to latency reductions of approximately 14.5\% and 13.3\%.
This confirms that maintaining hierarchical Morton order effectively removes sorting overhead and improves overall runtime efficiency.

All runtimes reported in this supplementary material were measured on a machine equipped with an NVIDIA GeForce RTX 3090 GPU and an Intel Core i9-9900K CPU with 64 GB of RAM.

\begin{table}[!h]
\caption{Runtime comparison between ELiC with explicit sorting and ELiC with Morton-order-preserving hierarchy.}
\label{tab:morton_runtime2}
\centering
\scriptsize
\begin{tabular}{lcccccc}
\hline
                & \multicolumn{2}{c}{\textbf{\begin{tabular}[c]{@{}c@{}}ELiC with\\ Explicit Sorting\end{tabular}}} & \multicolumn{2}{c}{\textbf{\begin{tabular}[c]{@{}c@{}}ELiC with \\ Morton-Order\end{tabular}}} & \multicolumn{2}{c}{\textbf{\begin{tabular}[c]{@{}c@{}}Latency\\ Reduction\end{tabular}}} \\
                & \textbf{enc (s)}                                & \textbf{dec (s)}                                & \textbf{enc (s)}                               & \textbf{dec (s)}                              & \textbf{enc (s)}                            & \textbf{dec (s)}                           \\ \hline
\textbf{16 bit} & 0.192                                           & 0.186                                           & 0.172                                          & 0.157                                         & -0.020                                      & -0.029                                     \\
\textbf{15 bit} & 0.171                                           & 0.157                                           & 0.145                                          & 0.135                                         & -0.026                                      & -0.022                                     \\
\textbf{14 bit} & 0.143                                           & 0.126                                           & 0.121                                          & 0.112                                         & -0.022                                      & -0.014                                     \\
\textbf{13 bit} & 0.113                                           & 0.097                                           & 0.095                                          & 0.086                                         & -0.018                                      & -0.011                                     \\
\textbf{12 bit} & 0.091                                           & 0.072                                           & 0.074                                          & 0.063                                         & -0.017                                      & -0.009                                     \\ \hline
\textbf{Avg.}   & 0.142                                           & 0.128                                          & 0.121                                          & 0.111                                         & -0.021                                      & -0.017                                     \\ \hline
\end{tabular}
\end{table}

\section{Asynchronous Arithmetic Encoding for Latency Reduction}
In the ELiC pipeline, the neural network inference and arithmetic encoding operate sequentially. 
This means the CPU must wait idly for the GPU to finish the neural network forward pass before the arithmetic encoding can begin, and conversely, the GPU remains idle while the CPU performs the encoding operation. 
To improve runtime efficiency, we implement an asynchronous pipeline that overlaps these two tasks on separate compute threads.

Specifically, the stage-wise occupancy probabilities $\{\mathbf{P}_{s}^{(b)}\}_{s{=}1,2}$ at bit-depth $b$ are emitted to a CPU buffer immediately upon computation.
While the GPU advances to bit-depth $b+1$, the CPU arithmetic-encodes the symbols from level $b$.
This producer-consumer pipeline overlaps neural network inference with entropy coding, reducing end-to-end encoding time without changing the bitstream or compression efficiency.

\begin{figure}[h!]
  \centering
  \includegraphics[width=\linewidth]{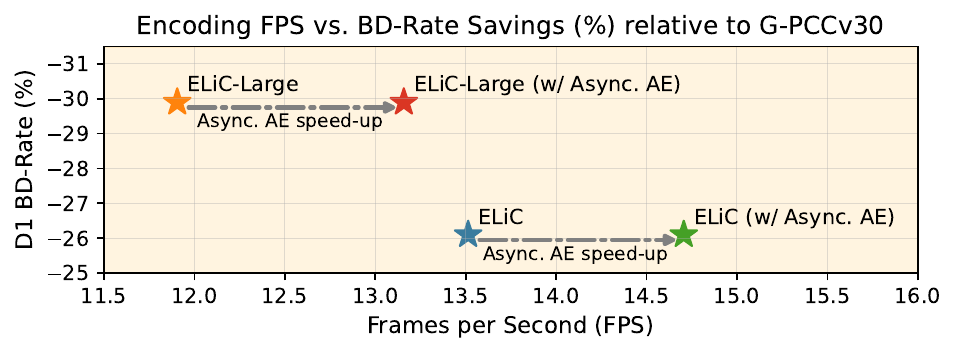}
  \caption{
    Speed-up achieved by applying asynchronous arithmetic encoding to ELiC.
  }
  \label{fig:async}
\end{figure}

Table~\ref{tab:async} compares end-to-end encoding latency with and without asynchronous arithmetic encoding. 
In our measurements, this asynchronous execution achieves an average reduction of 12\% in end-to-end encoding latency.
As shown in Fig.~\ref{fig:async}, applying asynchronous arithmetic encoding increases end-to-end encoding throughput by more than 1~FPS for both ELiC and ELiC-Large on 12-bit-depth LiDAR geometries.
This demonstrates that effective scheduling of the encoding pipeline, rather than modifying the model architecture, can further reduce latency.

\textit{Note that the runtime numbers reported in the main paper do not include this additional speed-up from asynchronous arithmetic encoding.}

\begin{table}[!h]
\caption{Encoding latency reduction (seconds per frame) achieved by applying asynchronous arithmetic encoding to ELiC.}
\label{tab:async}
\centering
\scriptsize
\begin{tabular}{lcccc}
\hline
\multicolumn{1}{c}{} & \textbf{ELiC} & \textbf{\begin{tabular}[c]{@{}c@{}}ELiC w/\\ Async. AE\end{tabular}} & \textbf{ELiC-Large} & \textbf{\begin{tabular}[c]{@{}c@{}}ELiC-Large w/\\ Async. AE\end{tabular}} \\ \hline
\textbf{16 bit}      & 0.172         & 0.145                                                                & 0.198               & 0.170                                                                      \\
\textbf{15 bit}      & 0.145         & 0.125                                                                & 0.162               & 0.141                                                                      \\
\textbf{14 bit}      & 0.121         & 0.105                                                                & 0.140               & 0.117                                                                      \\
\textbf{13 bit}      & 0.095         & 0.086                                                                & 0.105               & 0.094                                                                      \\
\textbf{12 bit}      & 0.074         & 0.068                                                                & 0.084               & 0.076                                                                      \\ \hline
\textbf{Avg.}        & 0.121         & 0.106                                                                & 0.138               & 0.120                                                                      \\ \hline
\end{tabular}
\end{table}

\section{Model Size and Parameter Efficiency}

Table~\ref{tab:params} reports the number of model parameters and the relative model size of RENO, RENO-Large, ELiC w/o BoE, ELiC, and ELiC-Large, all normalized by the RENO baseline.
The reported BD-Rate values are computed on the SemanticKITTI dataset in terms of D1-PSNR, as RENO-Large did not exhibit competitive performance on the Ford dataset, and SemanticKITTI was therefore used for comparison.
Compared to RENO, which is a very small model with 0.29M parameters, ELiC w/o BoE remains very small—incurring only a 17\% increase in the number of parameters—yet achieves a 6.61\% BD-Rate saving.
ELiC and ELiC-Large fall into the small-to-medium model regime with 2.04M and 8.04M parameters, respectively, and achieve D1-PSNR BD-Rate savings of 10.47\% and 15.09\% over RENO.
RENO-Large reduces the bitrate by 13.43\% relative to RENO, but its 20.55M parameters make it 2.56 times larger than ELiC-Large.
Overall, these results indicate that the ELiC family uses model capacity more efficiently than RENO while achieving better compression performance.

\begin{table}[h!]
\caption{Number of model parameters and relative size.}
\label{tab:params}
\centering
\scriptsize
\begin{tabular}{lrrr}
\hline
\textbf{Model}              & \multicolumn{1}{c}{\textbf{\begin{tabular}[c]{@{}c@{}}\# Parameters \\ (Millions)\end{tabular}}} & \multicolumn{1}{c}{\textbf{\begin{tabular}[c]{@{}c@{}}Relative \\ Ratio\end{tabular}}} & \multicolumn{1}{c}{\textbf{\begin{tabular}[c]{@{}c@{}}D1 BD-Rate (\%)\\ (SemanticKITTI, \\ vs. RENO)\end{tabular}}} \\ \hline
\textbf{RENO}               & 0.29 M                                                                                           & 1.00$\times$                                                                           & -                                                                                                                   \\
\textbf{RENO-Large}         & 20.55 M                                                                                          & 71.21$\times$                                                                          & -13.43\%                                                                                                            \\ \hdashline
\textbf{ELiC w/o BoE}       & 0.34 M                                                                                           & 1.17$\times$                                                                           & -6.61\%                                                                                                             \\
\textbf{ELiC ($K=5$)}       & 2.02 M                                                                                           & 6.99$\times$                                                                           & -10.53\%                                                                                                            \\
\textbf{ELiC-Large ($K=5$)} & 8.04 M                                                                                           & 27.86$\times$                                                                          & -15.63\%                                                                                                            \\ \hline
\end{tabular}
\end{table}

\section{Rate-Distortion Curves}
\begin{figure}[!t]
  \centering
  \begin{subfigure}[t]{0.9\linewidth}
    \centering
    \includegraphics[width=\linewidth]{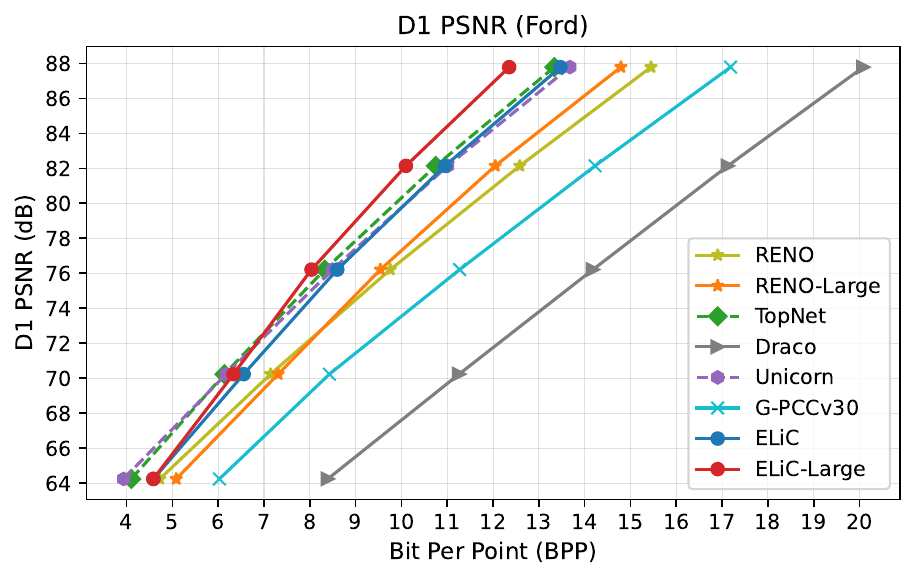}
    \caption{BPP vs. D1 PSNR (Ford)}
  \end{subfigure}
  
  \begin{subfigure}[t]{0.9\linewidth}
    \centering
    \includegraphics[width=\linewidth]{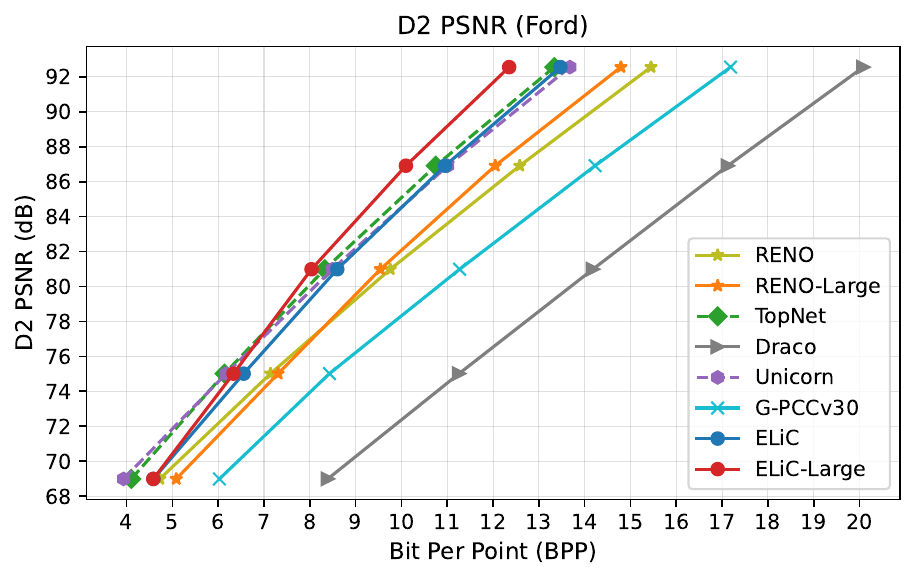}
    \caption{BPP vs. D2 PSNR (Ford)}
  \end{subfigure}
  
  \begin{subfigure}[t]{0.9\linewidth}
    \centering
    \includegraphics[width=\linewidth]{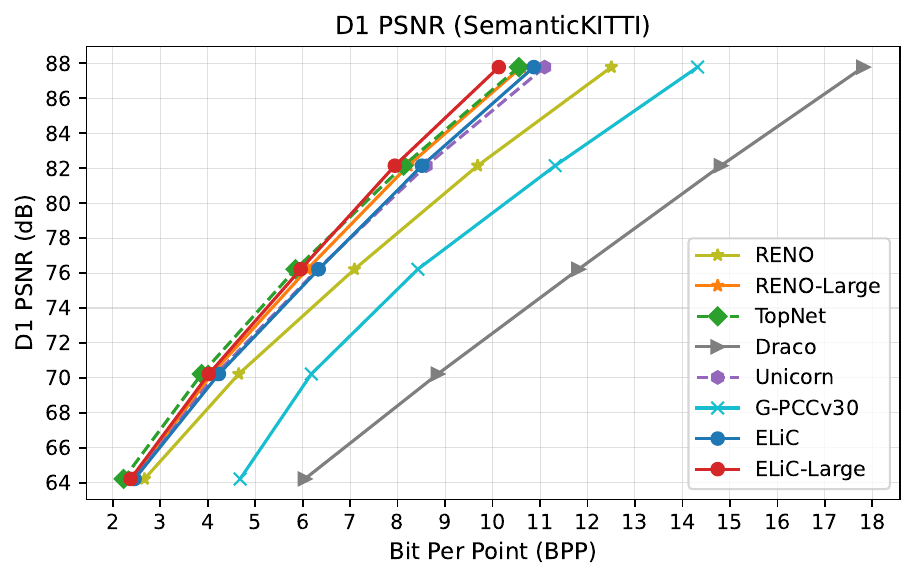}
    \caption{BPP vs. D1 PSNR (SemanticKITTI)}
  \end{subfigure}
  
  \begin{subfigure}[t]{0.9\linewidth}
    \centering
    \includegraphics[width=\linewidth]{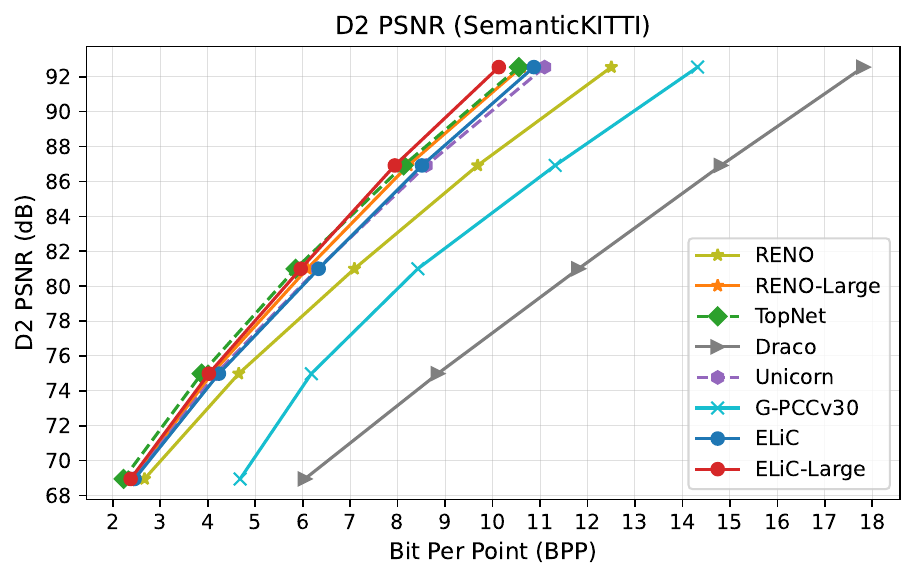}
    \caption{BPP vs. D2 PSNR (SemanticKITTI)}
  \end{subfigure}
  \caption{Rate-distortion curves from Table~\ref{tab:BDBR} in Sec.~\ref{sec:compres}}
  \label{fig:rdcurves}
\end{figure}

Fig.~\ref{fig:rdcurves} shows rate-distortion (R-D) curves derived from Table~\ref{tab:BDBR} in the main paper (Sec.~\ref{sec:compres}), used for the compression efficiency comparison.
Each curve contains five points representing the rate-distortion performance for input LiDAR geometries at 16, 15, 14, 13, and 12 bit-depths.

ELiC markedly improves R-D performance over the RENO baseline, and ELiC-Large achieves compression performance comparable to Unicorn and TopNet. 
Referencing Sec.~\ref{sec:runtime}, RENO-Large, Unicorn, and TopNet operate at least 3$\times$ higher time complexity than ELiC variants for both encoding and decoding, indicating that ELiC offers a substantially better compression-runtime trade-off and stronger practical deployability.

At the 16- and 15-bit coordinate inputs, the ELiC variants exhibit notably superior compression performance.
However, at 13- and 12-bit coordinate inputs, a slight degradation is observed.
This trend likely stems from the design of the training objective in Eq.~\ref{eq:train-loss}, which minimizes the total bit count for the entire hierarchy.
As a result, the model tends to allocate more capacity to support the highly sparse upper bit-depths, where a larger number of bits is required.
Future work should explore strategies to balance optimization across all bit-depth levels for more uniformly strong performance.

\section{Additional Bit-Depth-Wise Results}
\label{sec:add_results}

\begin{figure*}[!t]
\centering

\begin{minipage}[b]{0.05\textwidth}\centering
  \rotatebox{90}{\footnotesize \qquad\textbf{15 bit-depth level}}
\end{minipage}\hspace{0.01\textwidth}
\begin{subfigure}[b]{0.3\textwidth}
  \includegraphics[width=\textwidth]{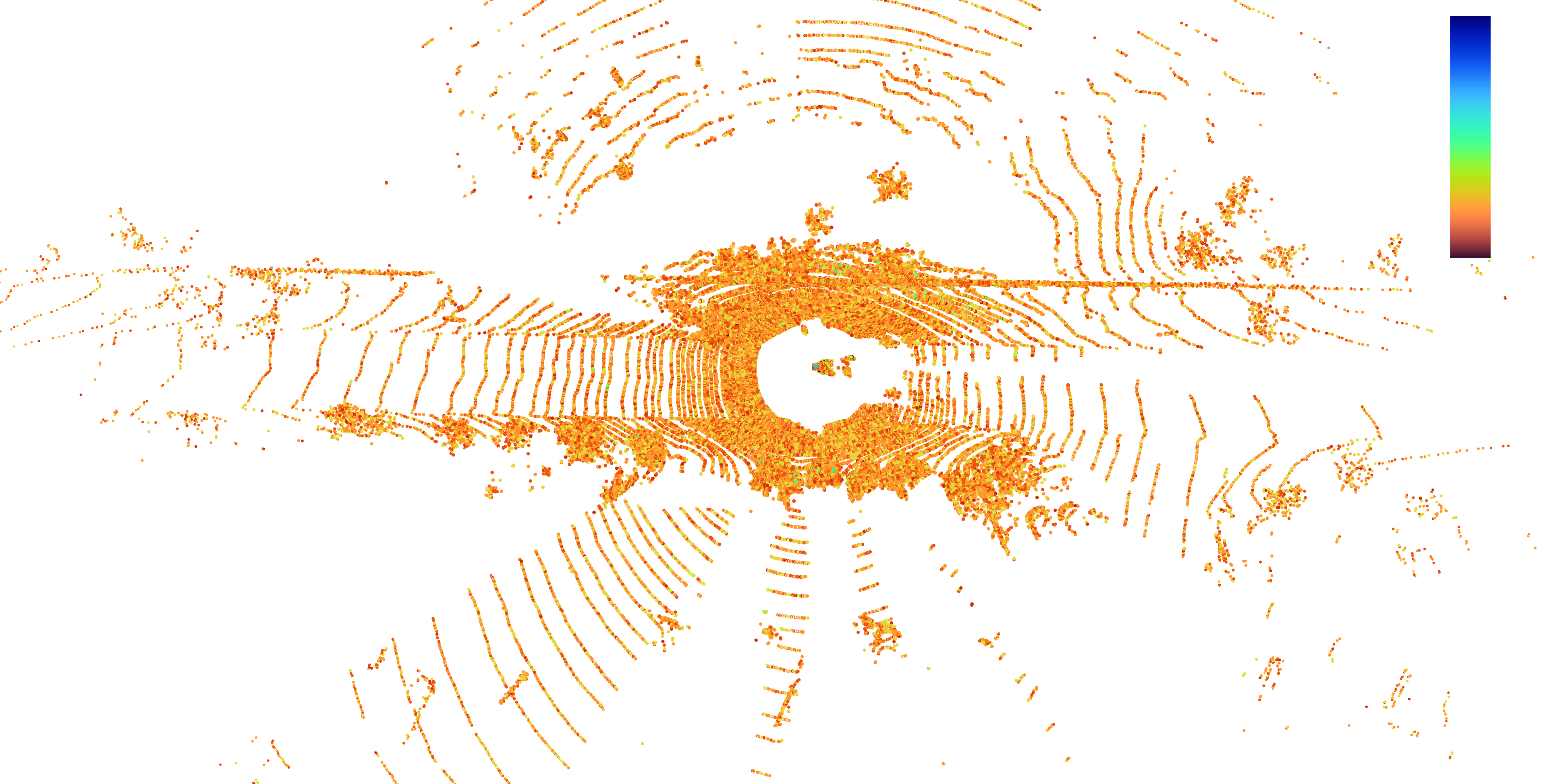}
  \caption{RENO, 248.54 KBits}
\end{subfigure}\hspace{0.01\textwidth}
\begin{subfigure}[b]{0.3\textwidth}
  \includegraphics[width=\textwidth]{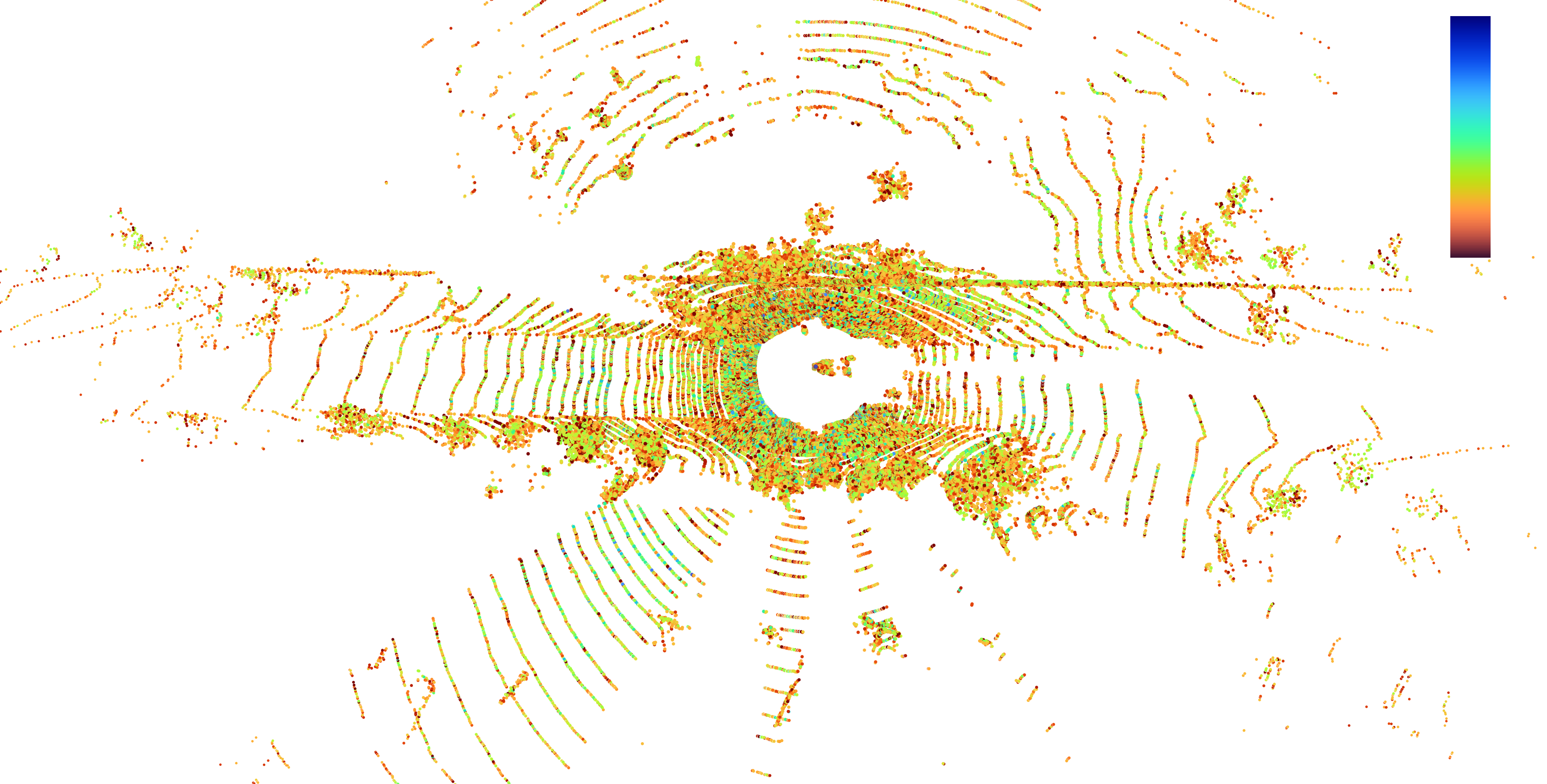}
  \caption{ELiC, 230.32 KBits}
\end{subfigure}\hspace{0.01\textwidth}
\begin{subfigure}[b]{0.3\textwidth}
  \includegraphics[width=\textwidth]{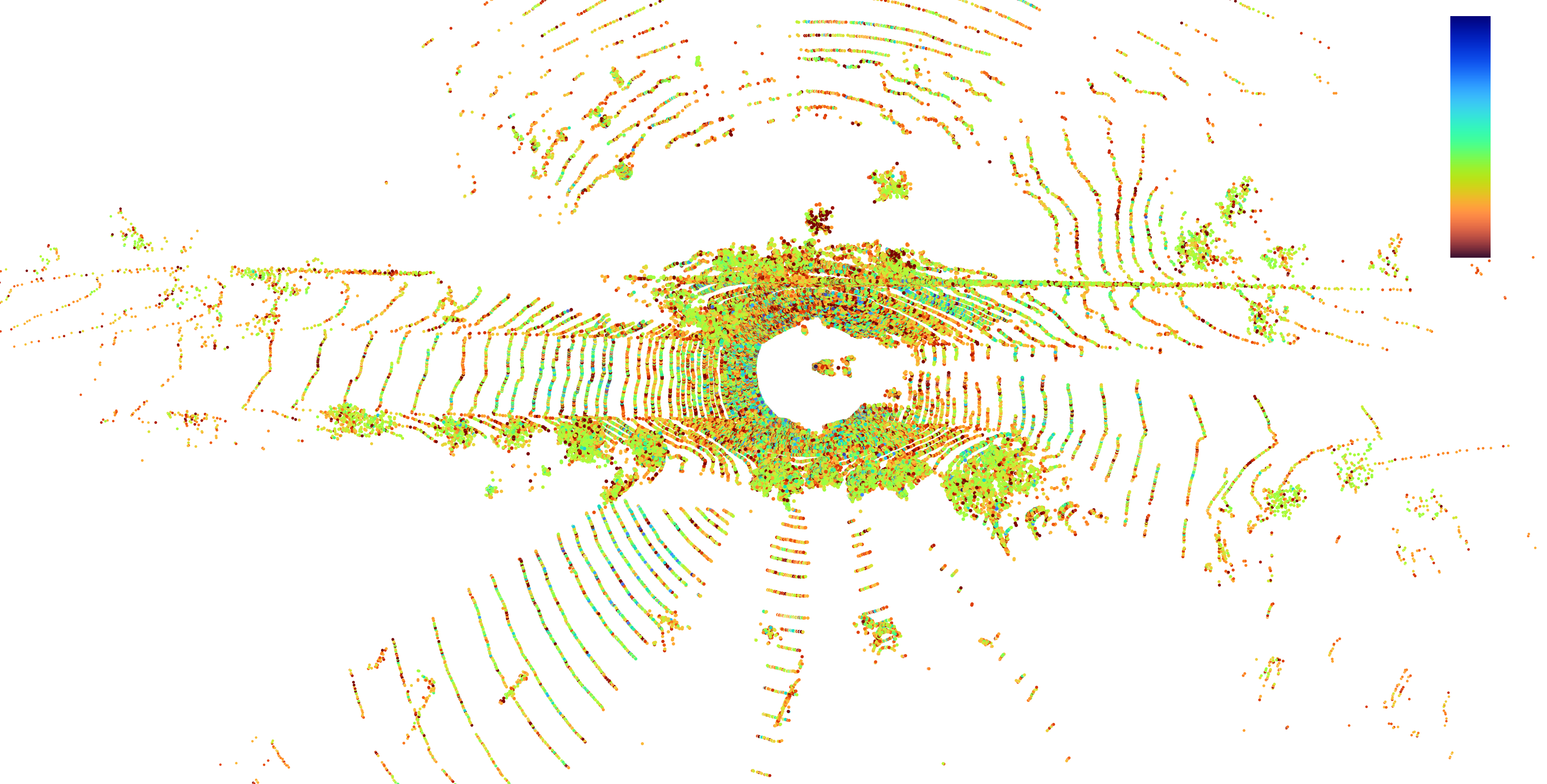}
  \caption{ELiC-Large, 219.43 KBits}
\end{subfigure}

\vspace{0.015\textwidth}

\begin{minipage}[b]{0.05\textwidth}\centering
  \rotatebox{90}{\footnotesize \qquad\textbf{13 bit-depth level}}
\end{minipage}\hspace{0.01\textwidth}
\begin{subfigure}[b]{0.3\textwidth}
  \includegraphics[width=\textwidth]{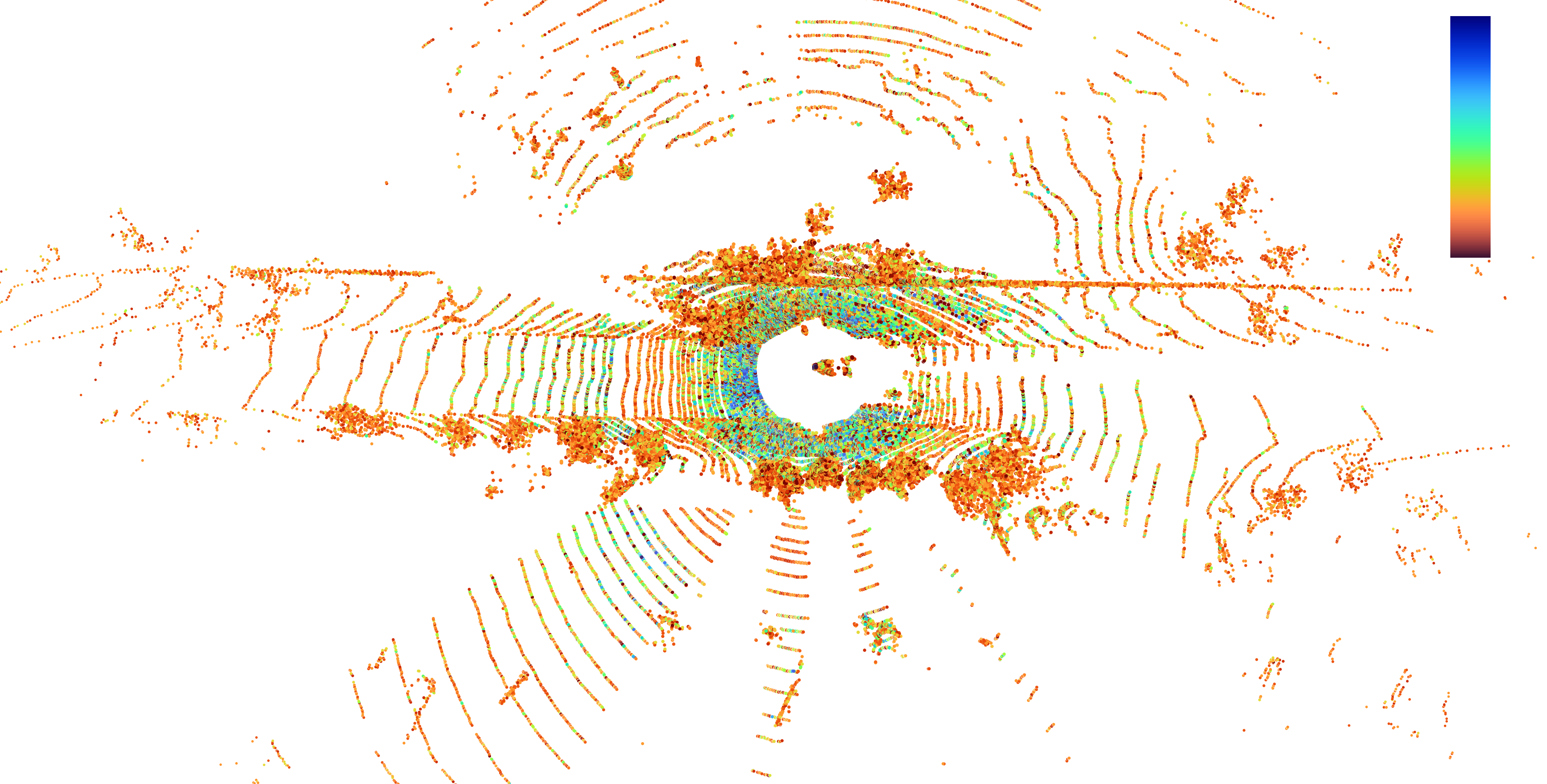}
  \caption{RENO, 240.55 KBits}
\end{subfigure}\hspace{0.01\textwidth}
\begin{subfigure}[b]{0.3\textwidth}
  \includegraphics[width=\textwidth]{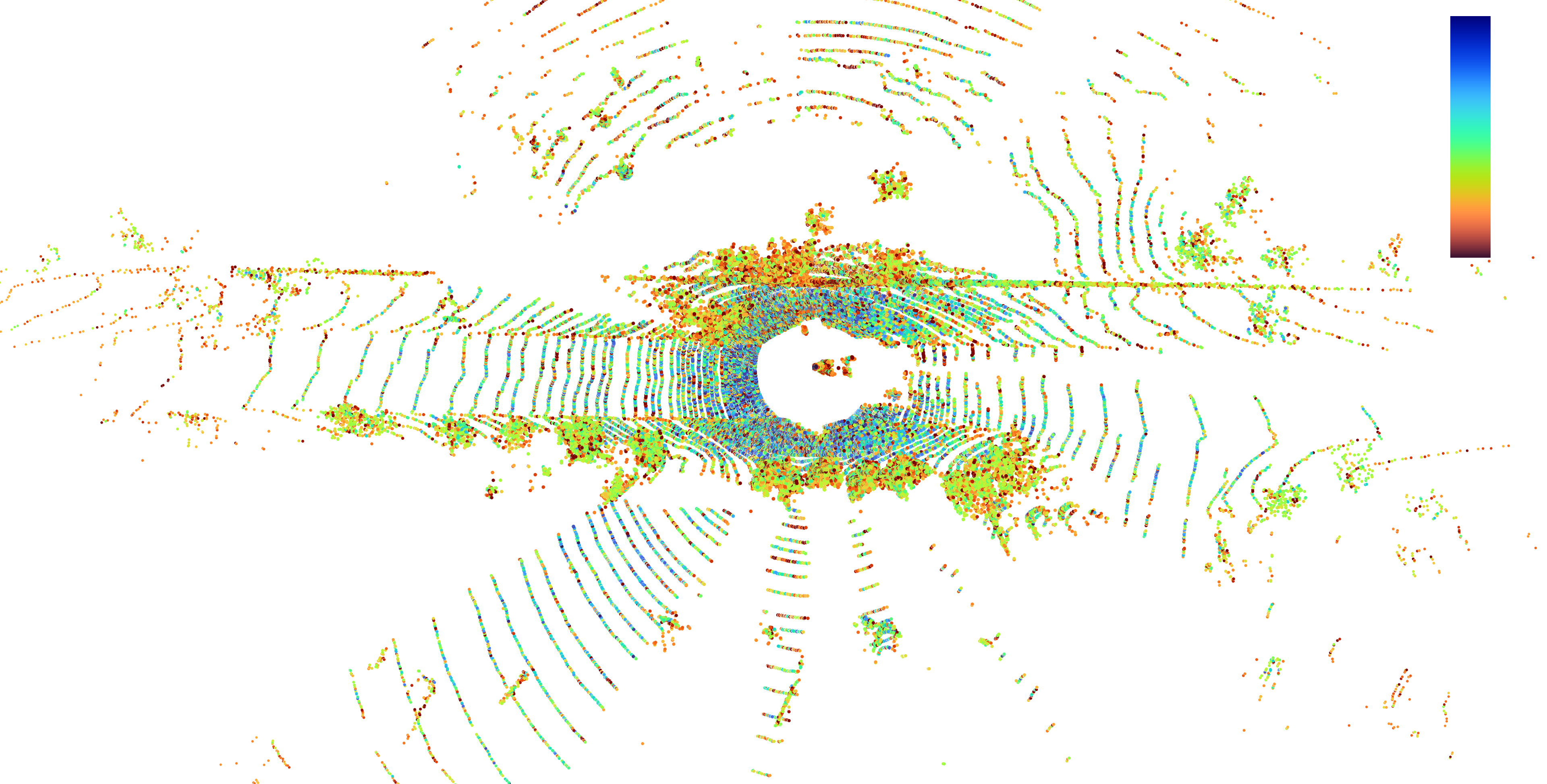}
  \caption{ELiC, 199.79 KBits}
\end{subfigure}\hspace{0.01\textwidth}
\begin{subfigure}[b]{0.3\textwidth}
  \includegraphics[width=\textwidth]{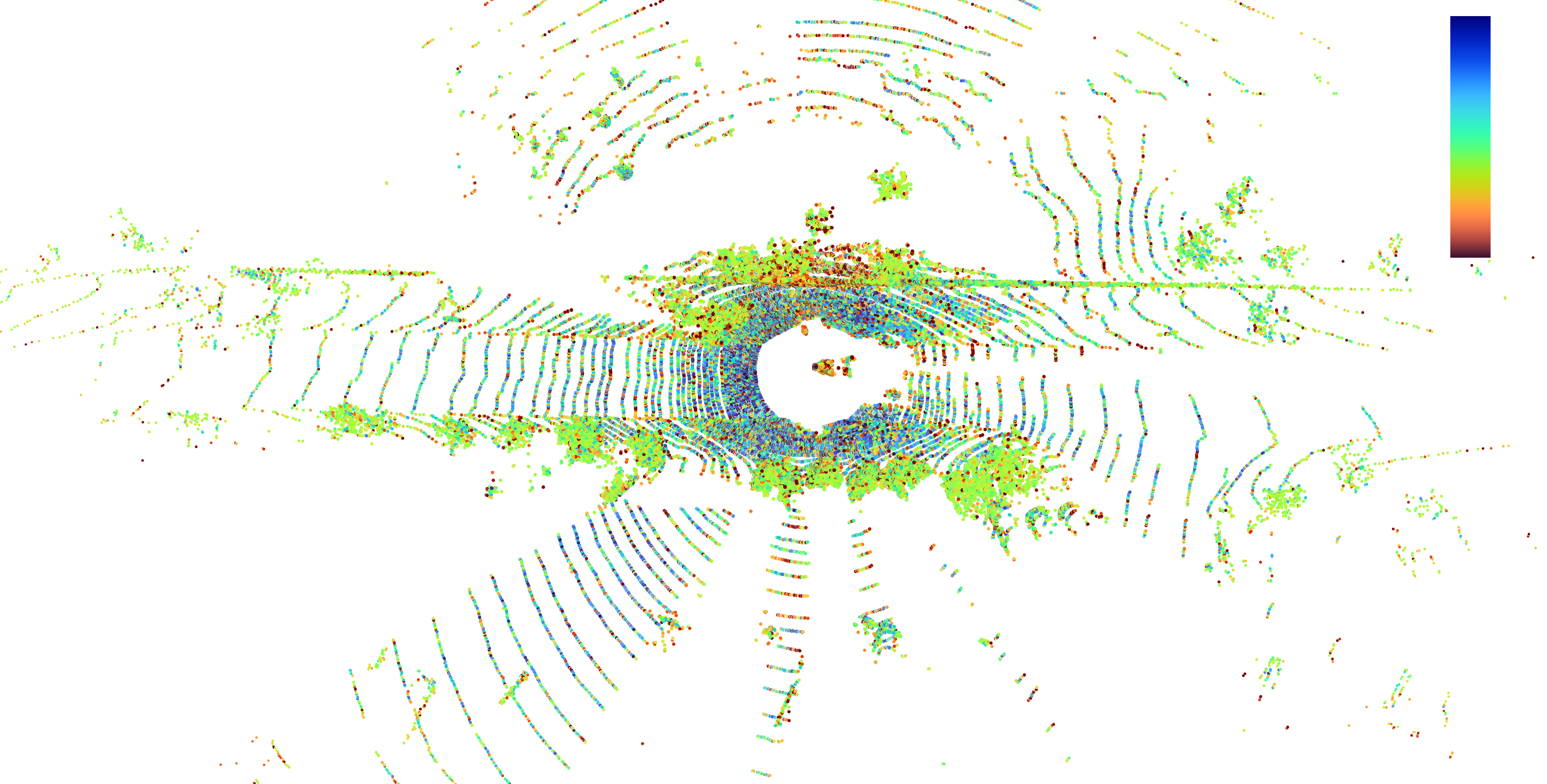}
  \caption{ELiC-Large, 184.38 KBits}
\end{subfigure}

\begin{minipage}[b]{0.05\textwidth}\centering
  \rotatebox{90}{\footnotesize \qquad\textbf{11 bit-depth level}}
\end{minipage}\hspace{0.01\textwidth}
\begin{subfigure}[b]{0.3\textwidth}
  \includegraphics[width=\textwidth]{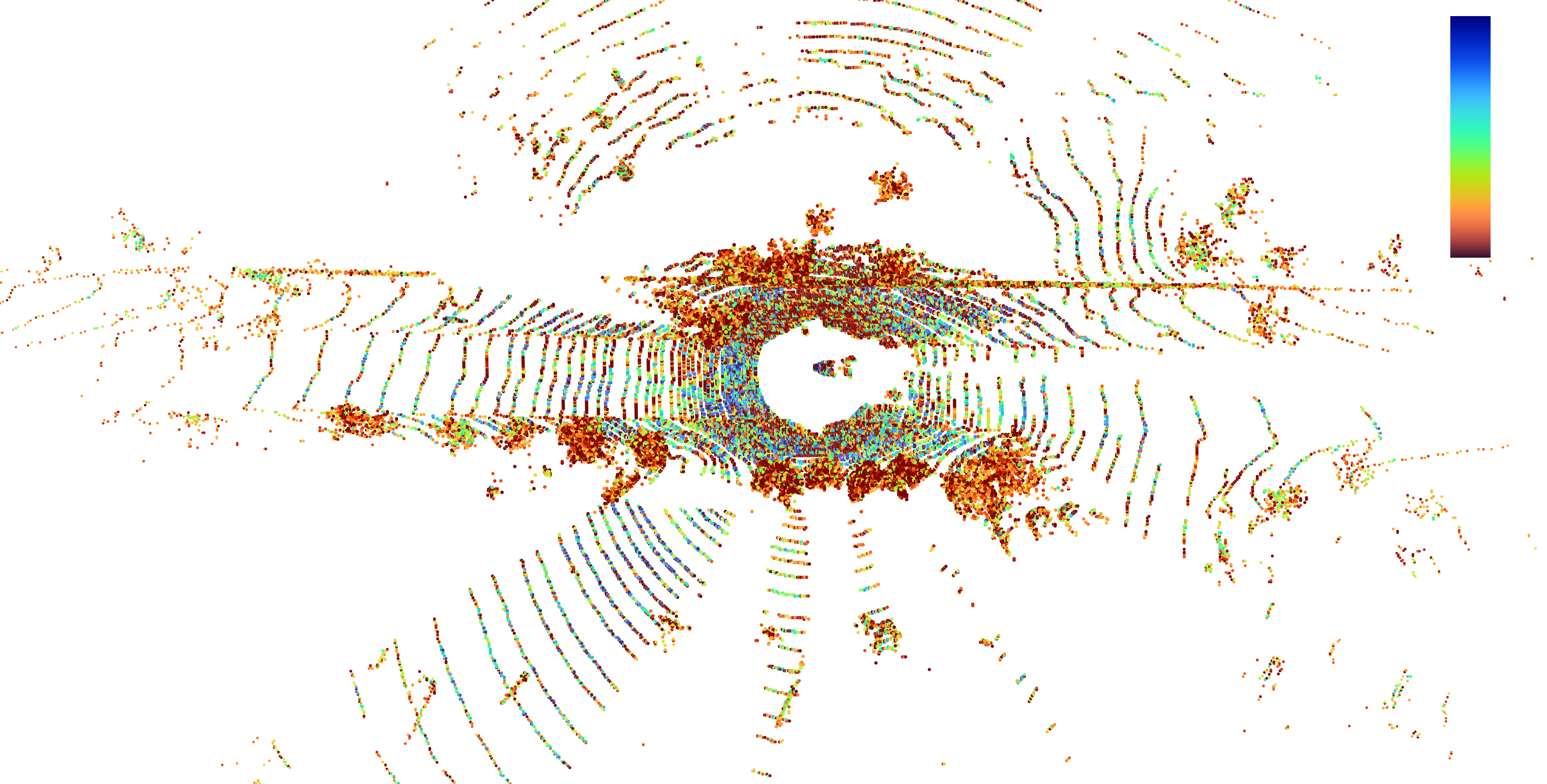}
  \caption{RENO, 192.51 KBits}
\end{subfigure}\hspace{0.01\textwidth}
\begin{subfigure}[b]{0.3\textwidth}
  \includegraphics[width=\textwidth]{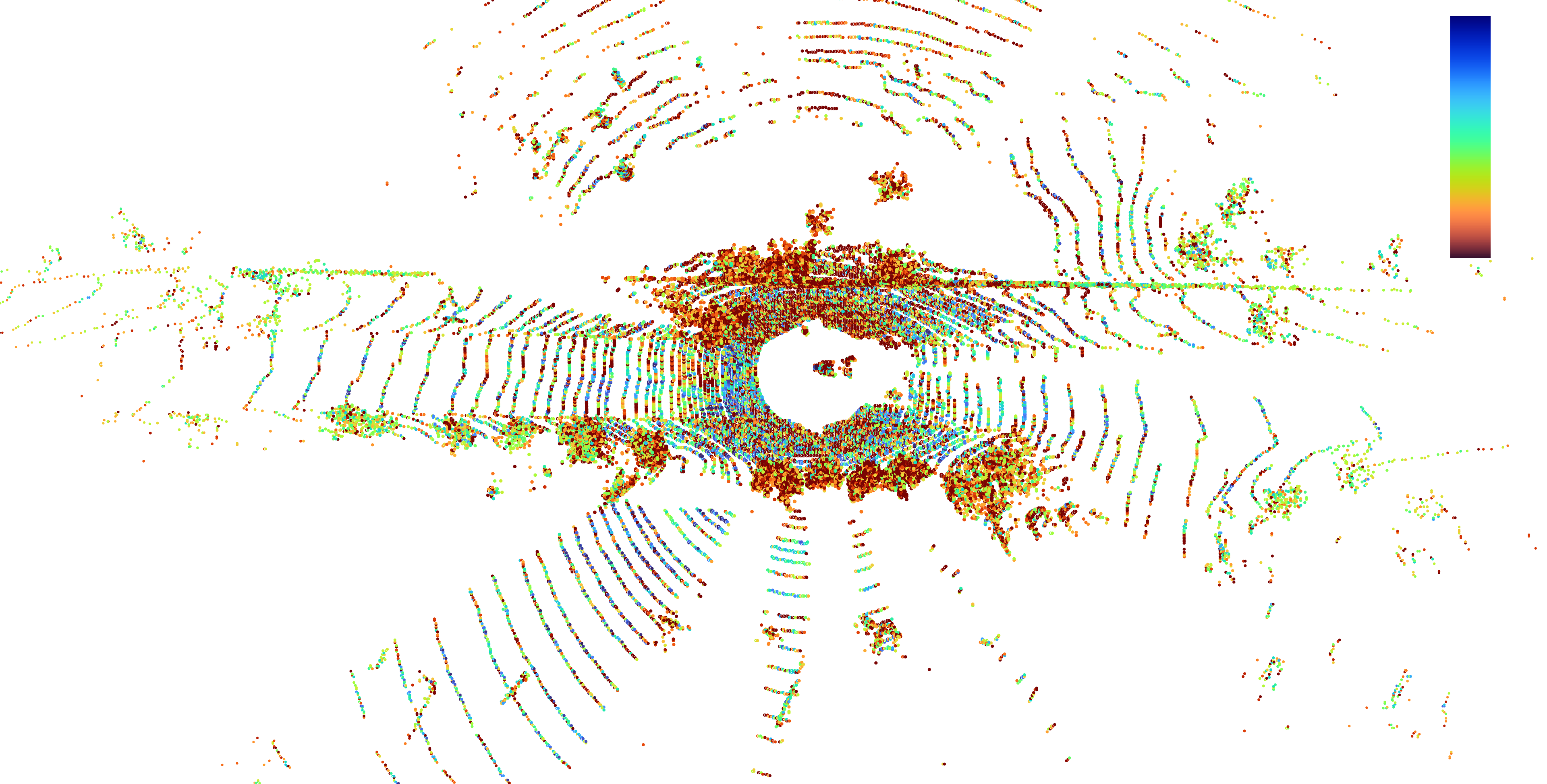}
  \caption{ELiC, 180.42 KBits}
\end{subfigure}\hspace{0.01\textwidth}
\begin{subfigure}[b]{0.3\textwidth}
  \includegraphics[width=\textwidth]{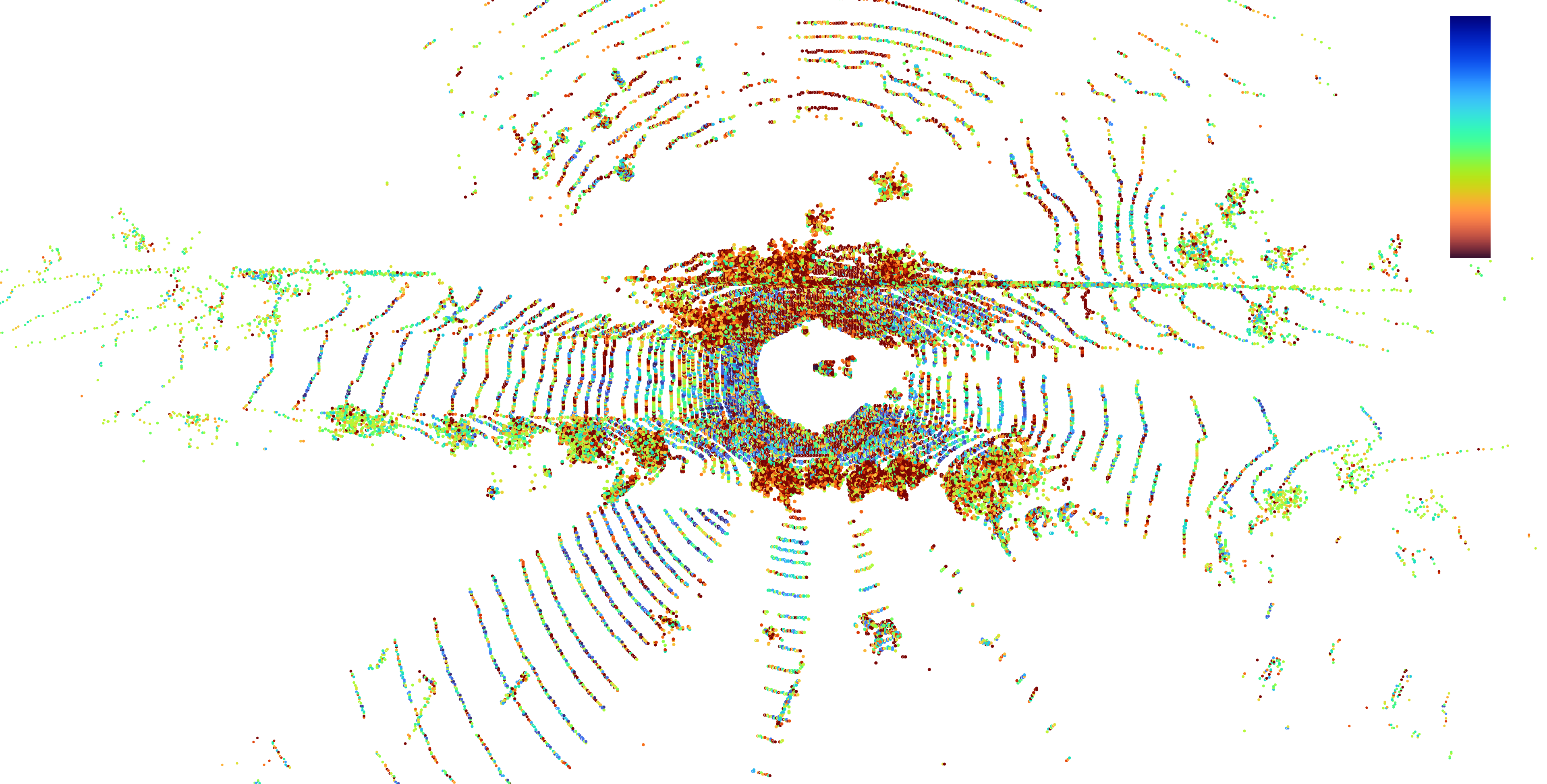}
  \caption{ELiC-Large, 171.35 KBits}
\end{subfigure}

\begin{minipage}[b]{0.05\textwidth}\centering
  \rotatebox{90}{\footnotesize \qquad\textbf{9 bit-depth level}}
\end{minipage}\hspace{0.01\textwidth}
\begin{subfigure}[b]{0.3\textwidth}
  \includegraphics[width=\textwidth]{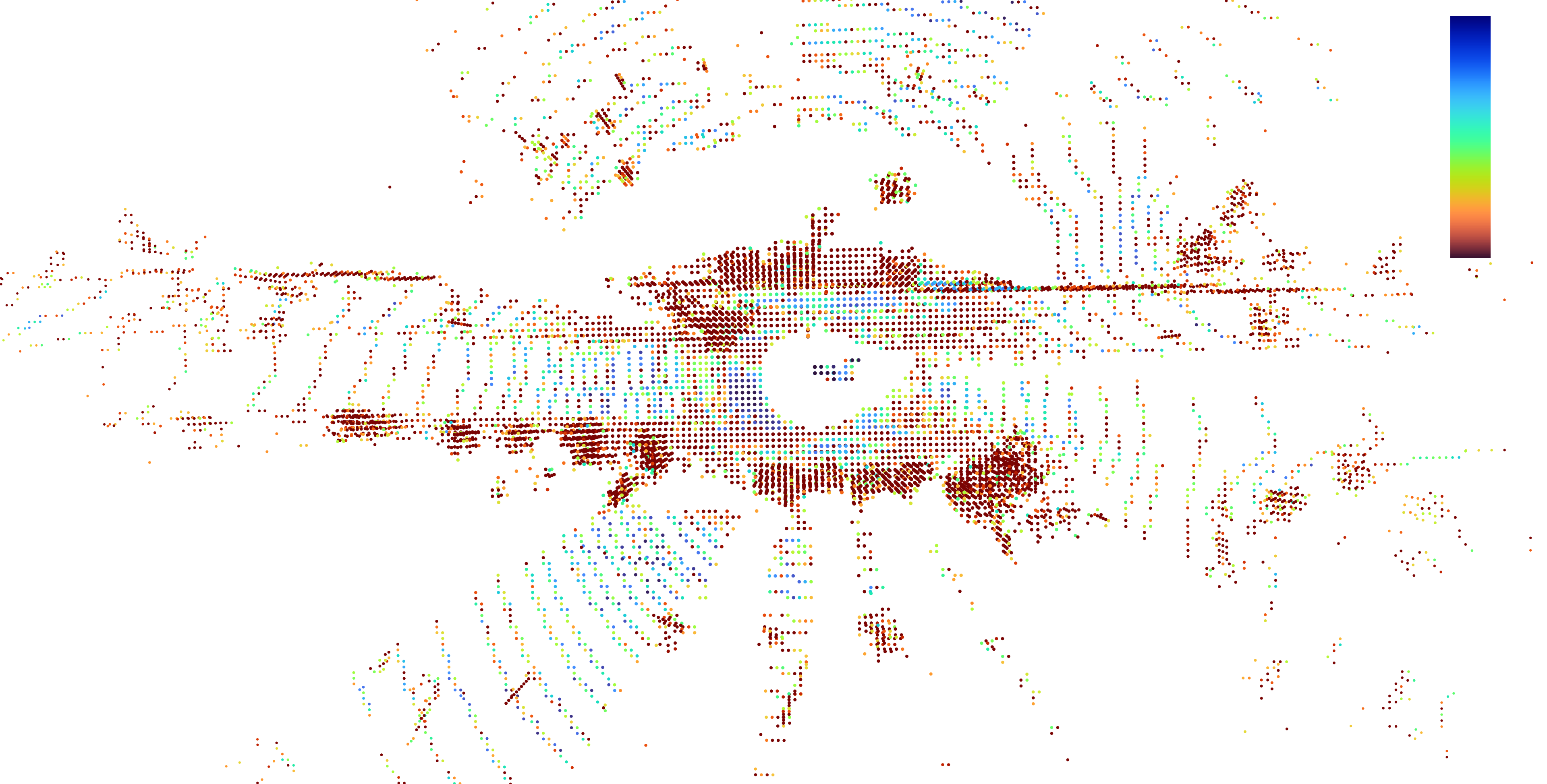}
  \caption{RENO, 50.63 KBits}
\end{subfigure}\hspace{0.01\textwidth}
\begin{subfigure}[b]{0.3\textwidth}
  \includegraphics[width=\textwidth]{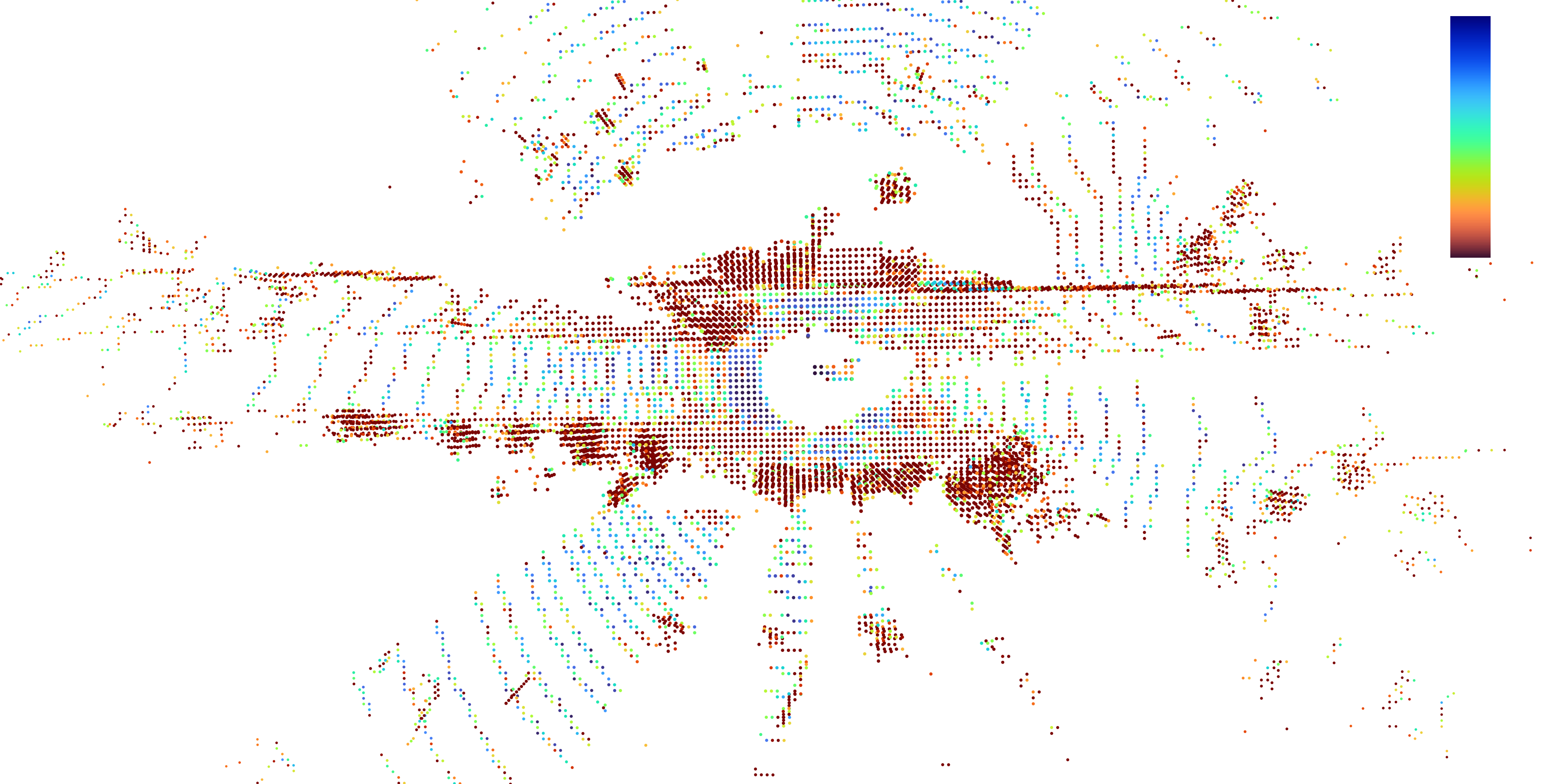}
  \caption{ELiC, 47.81 KBits}
\end{subfigure}\hspace{0.01\textwidth}
\begin{subfigure}[b]{0.3\textwidth}
  \includegraphics[width=\textwidth]{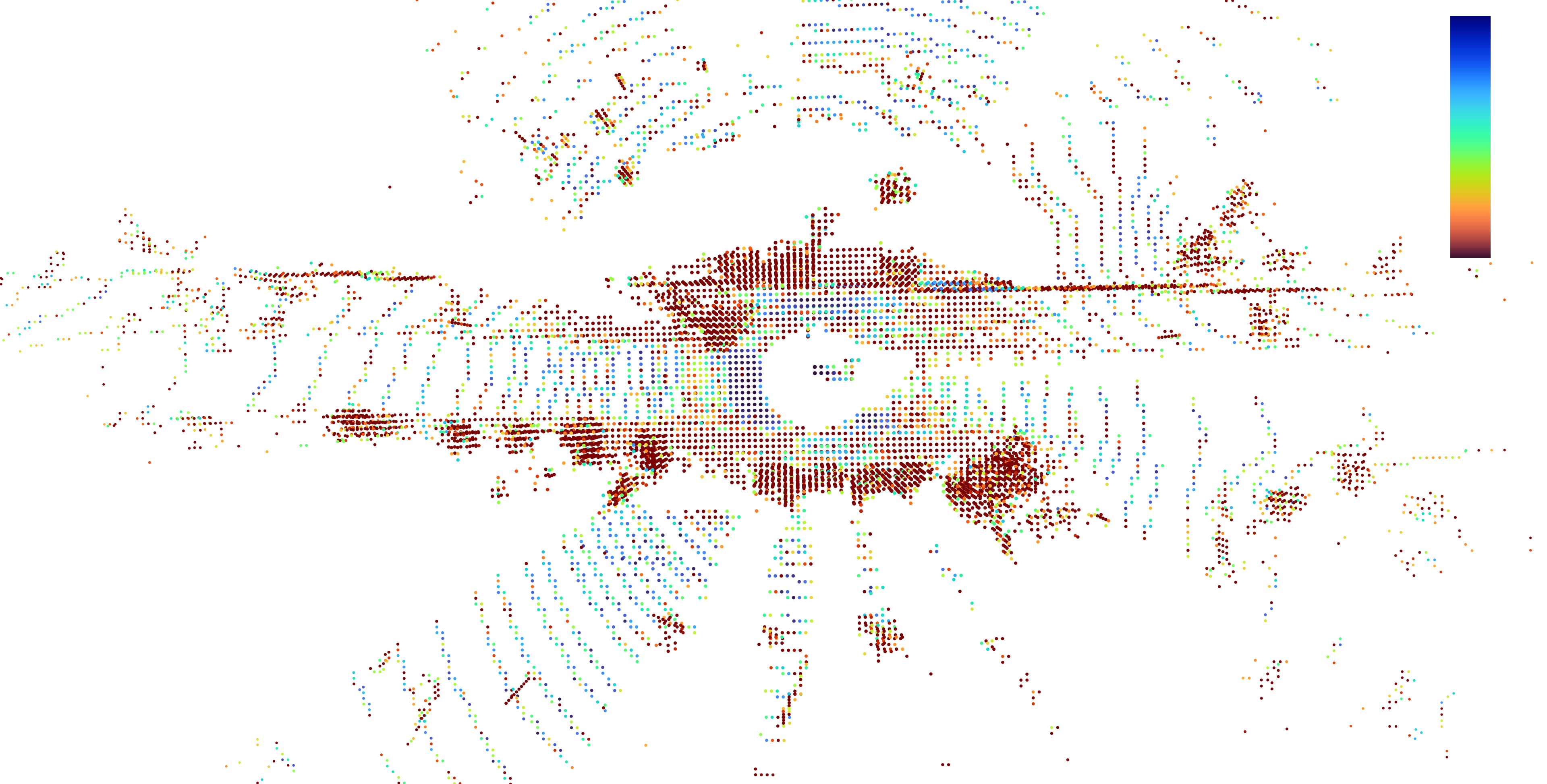}
  \caption{ELiC-Large, 46.46 KBits}
\end{subfigure}

\begin{minipage}[b]{0.05\textwidth}\centering
  \rotatebox{90}{\footnotesize \qquad\textbf{7 bit-depth level}}
\end{minipage}\hspace{0.01\textwidth}
\begin{subfigure}[b]{0.3\textwidth}
  \includegraphics[width=\textwidth]{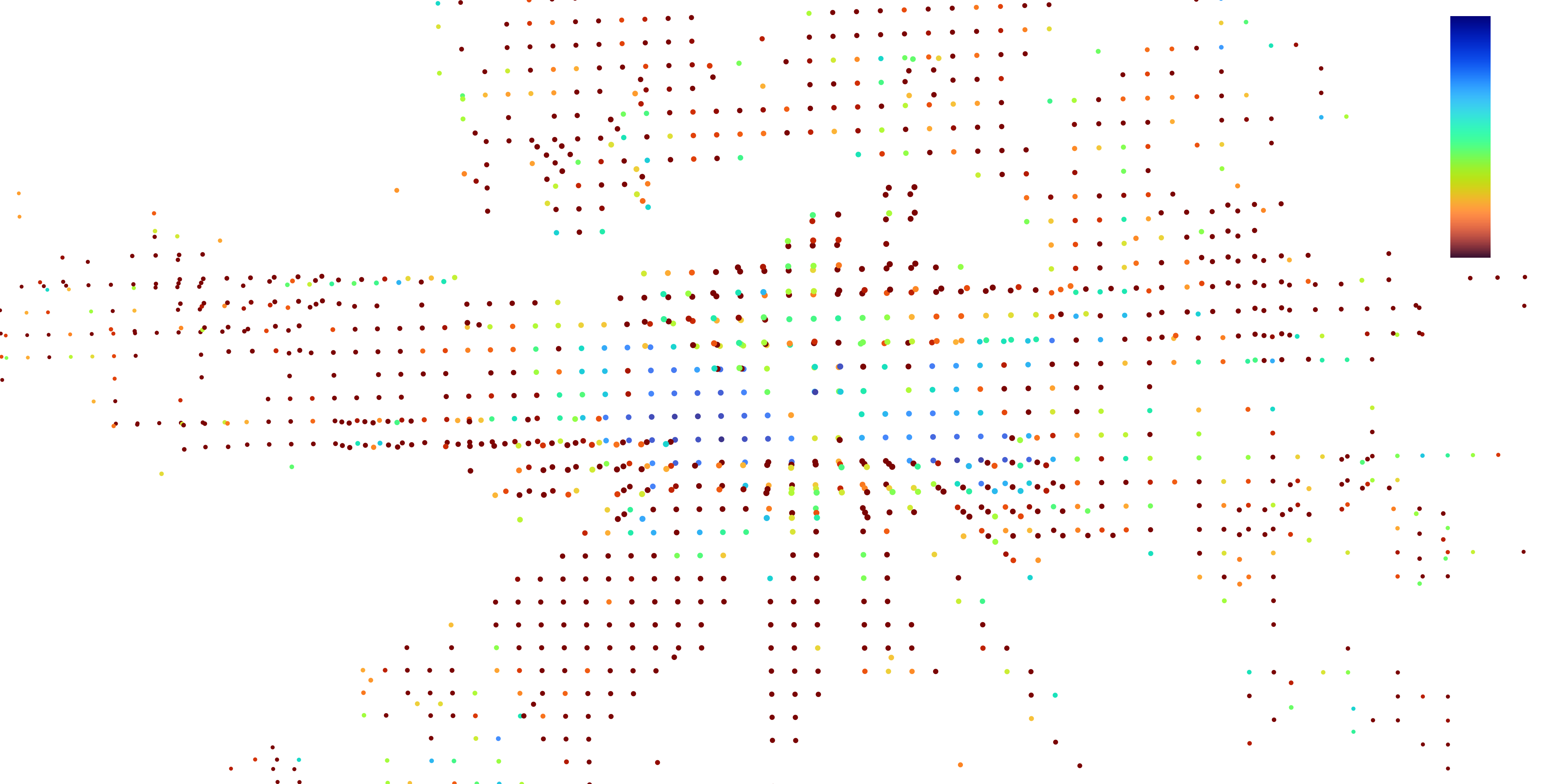}
  \caption{RENO, 9.34 KBits}
\end{subfigure}\hspace{0.01\textwidth}
\begin{subfigure}[b]{0.3\textwidth}
  \includegraphics[width=\textwidth]{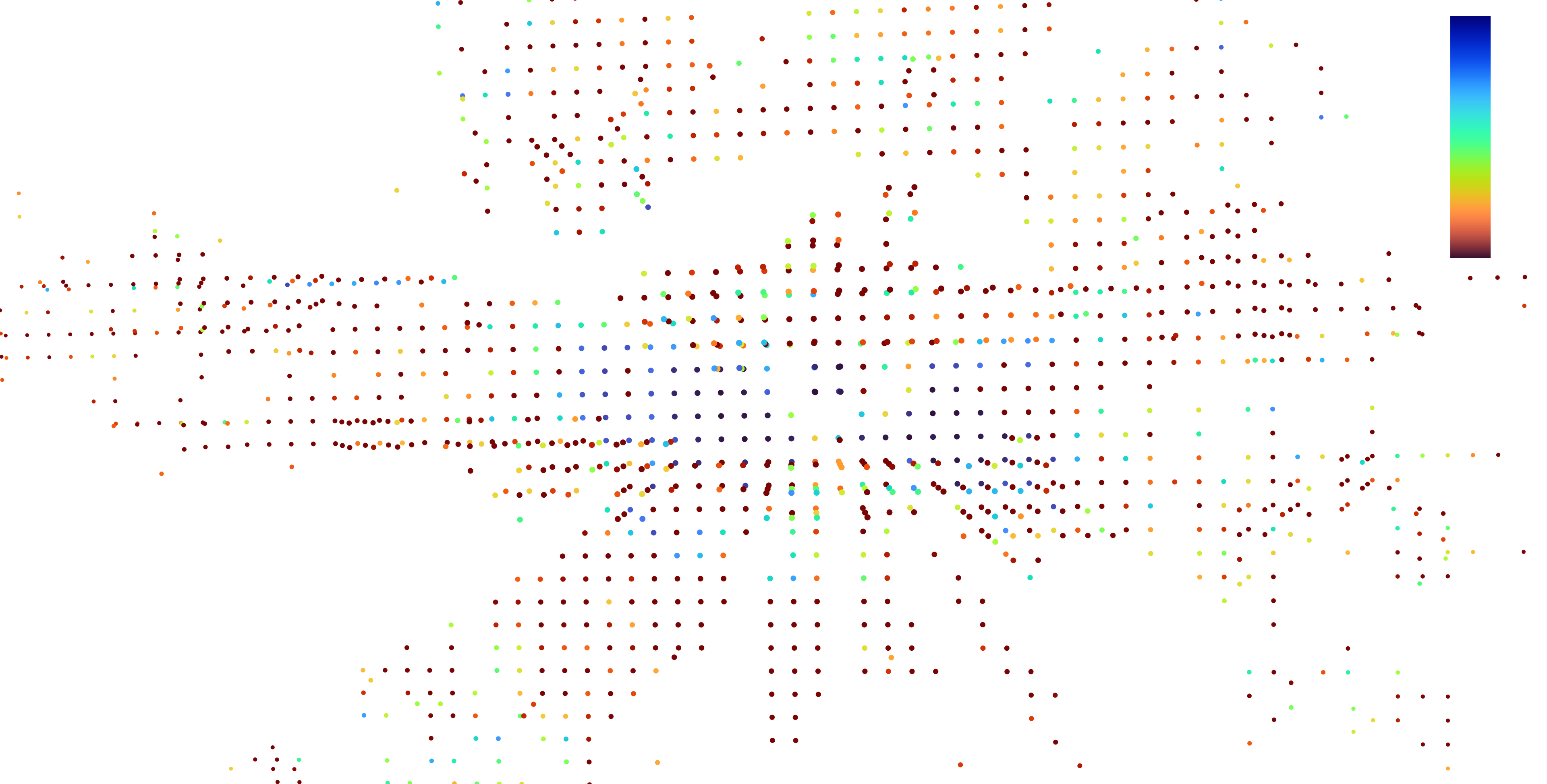}
  \caption{ELiC, 8.87 KBits}
\end{subfigure}\hspace{0.01\textwidth}
\begin{subfigure}[b]{0.3\textwidth}
  \includegraphics[width=\textwidth]{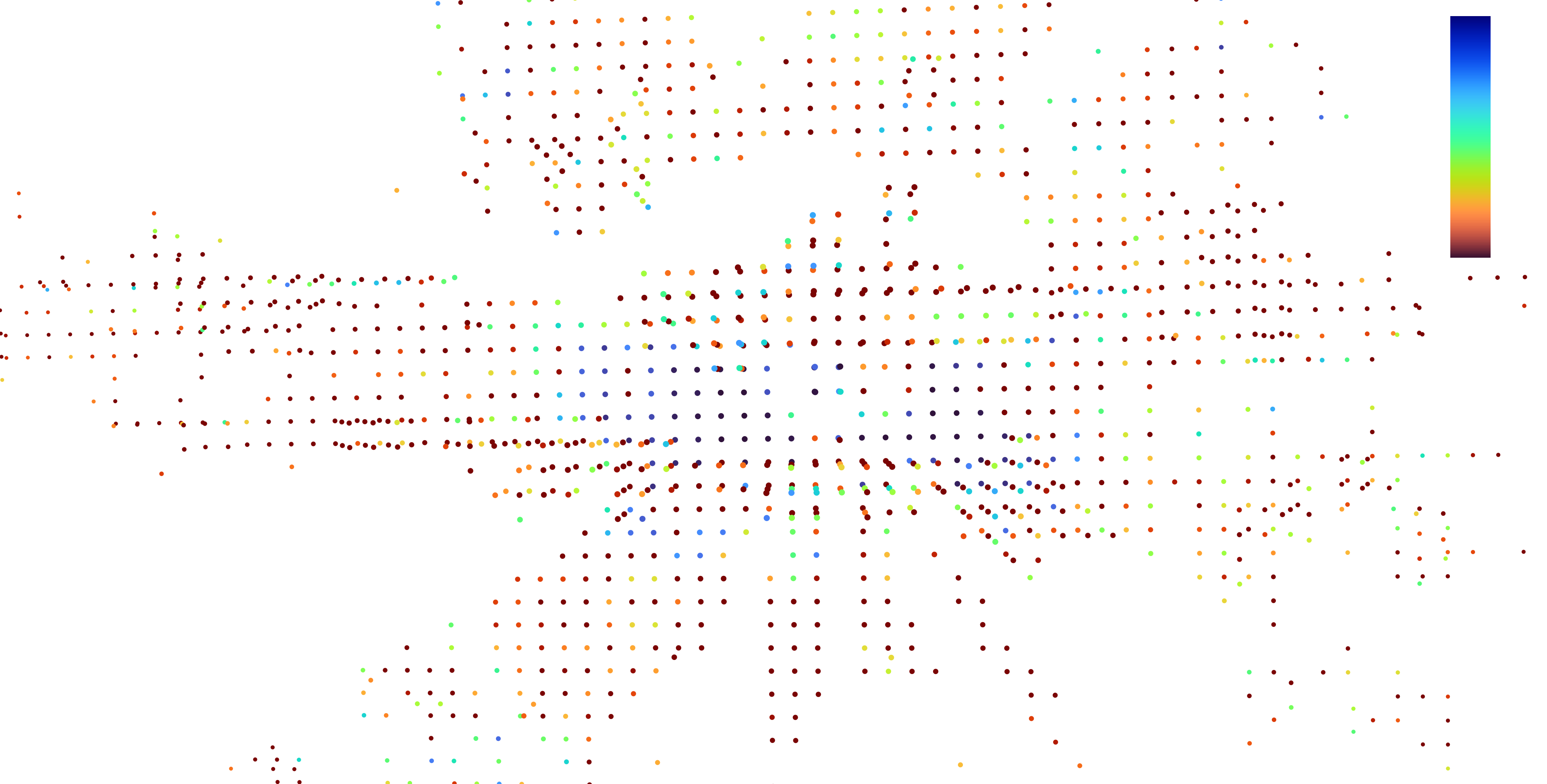}
  \caption{ELiC-Large, 8.86 KBits}
\end{subfigure}

\caption{Per-point bit allocation on the ``Ford\_03\_vox1mm-0200.ply" frame at the 15, 13, 11, and 9 bit-depth levels for RENO, ELiC, and ELiC-Large.
For each bit-depth level, the colormap is normalized using the minimum and maximum predicted bits among the three models.
The colormap ranges from lower predicted bits (blue) to higher predicted bits (dark orange) per point.}
\label{fig:bitalloc_supp}

\end{figure*}

Fig.~\ref{fig:bitalloc_supp} visualizes per-point bit cost when compressing a 16-bit LiDAR frame with RENO, ELiC, and ELiC-Large. 
The per-point bit cost at bit-depth level $b$ is the estimated code length from Eq.~\eqref{eq:train-loss}, i.e., for point $n$ it equals
$-\log_{2}\mathbf{P}_{s{=}1,k}^{(b)}\big(n,\mathbf{Q}_{s{=}1}^{(b)}(n)\big){-}\log_{2}\mathbf{P}_{s{=}2,k}^{(b)}\big(n,\mathbf{Q}_{s{=}2}^{(b)}(n)\big)$.
We show levels $b\in\{15,13,11,9,7\}$ and annotate the total bits for each level. 
RENO allocates similar totals at $b{=}13$ and $b{=}15$ (about $240.6$ vs.\ $248.5$ KBits), which indicates weak context exploitation in the sparse regime and limited occupancy prediction accuracy. 
ELiC and ELiC-Large yield lower bit costs at $b{=}13$ and $b{=}15$ bit-depth levels and maintain better compression performance in far-range areas.

Across all three models, the gap between ground and non-ground grows as bit-depth level decreases and density increases. 
Non-ground regions are complex and irregular, which aligns poorly with a fixed $4{+}4$ two-stage factorization. The bit grouping does not always match spatial correlation, and the best factorization likely depends on density and local structure.

At present, all BoE coding networks use the same $4{+}4$ split. 
A natural extension is to let BoE select not only the network parameters but also the symbol factorization, so that different occupancy distributions can choose a more suitable split per level.

